\documentclass[aps, prd,twocolumn,superscriptaddress,nofootinbib,preprintnumbers]{revtex4-1}
\usepackage{graphicx}
\usepackage{amsmath,amssymb}
\usepackage{hyperref}
\usepackage{braket}
\usepackage{subcaption}
\usepackage{float}
\usepackage[dvipsnames]{xcolor}
\usepackage{soul}
\usepackage{rotating}
\usepackage{multirow}
\usepackage{mathtools}
\usepackage{makecell}

\usepackage{txfonts}
\usepackage[margin=0.75in]{geometry}
\usepackage[normalem]{ulem}
\usepackage{graphicx}
\usepackage[justification=raggedright, singlelinecheck=false]{caption}
\usepackage[capitalise]{cleveref}

\newcommand{\be}{\begin{equation}}                                  
\newcommand{\ee}{\end{equation}}                                    
\newcommand{\ba}{\begin{eqnarray}}                                  
\newcommand{\ea}{\end{eqnarray}}                                    
                                          
\newcommand{\barr}{\begin{array}}                                    
\newcommand{\earr}{\end{array}}

\newcommand{\penn}{Department of Physics and Astronomy, University of Pennsylvania, Philadelphia, PA 19104, USA}
\newcommand{\perimeter}{Perimeter Institute for Theoretical Physics, 31 Caroline St North, Waterloo, ON N2L 2Y5, Canada}
\newcommand{\lorentz}{Lorentz Institute for Theoretical Physics, Leiden University, PO box 9506, 2300 RA Leiden, the Netherlands}
\newcommand{\leiden}{Leiden Observatory, Leiden University, PO Box 9513, 2300 RA Leiden, the Netherlands}

\begin{document}

\title[The kSZ to $P_{\rm mm}$ Emulator]{Inferring the Impacts of Baryonic Feedback from Kinetic Sunyaev-Zeldovich Cross-Correlations}

\author{Alex Lagu\"e}\affiliation{\penn}
\author{Mathew S. Madhavacheril}\affiliation{\penn}
\author{Josh Borrow}\affiliation{\penn}
\author{\\ Kendrick M. Smith}\affiliation{\perimeter} 
\author{Xinyi Chen}\affiliation{Center for Cosmology and AstroParticle Physics (CCAPP), The Ohio State University, Columbus, OH 43210, USA}
\author{Matthieu Schaller}\affiliation{\lorentz}\affiliation{\leiden}
\author{Joop Schaye} \affiliation{\leiden}

\begin{abstract}
The complex processes of baryonic feedback associated with galaxy evolution are still poorly understood, and their impact on the clustering of matter on small scales remains difficult to quantify. While many fitting functions and emulators exist to model the matter power spectrum, their input parameters are not directly observable. However, recent studies using hydrodynamical simulations have identified a promising correlation between the gas content of halos and changes to the matter power spectrum from feedback. Building on these findings, we create the first fully data-driven power spectrum emulator. We utilize the kinematic Sunyaev-Zeldovich (kSZ) effect, a secondary anisotropy in the cosmic microwave background, as a tracer of free electrons in and around halos. We train a neural network to learn the mapping between the suppression of the matter power spectrum and the shape of the kSZ power spectrum extracted with a radial velocity template. We train and validate our algorithm using the FLAMINGO suite of hydrodynamical simulations, which encompasses a wide range of feedback models. Our emulator can reconstruct the matter power spectrum at the sub-percent level for scales $k\leq 5\;h/$Mpc and $0.2\leq z \leq 1.25$ directly from the data. Our model is robust and retains percent-level accuracy even for feedback models and cosmological parameter values not seen during training (except in a few extreme cases drastically different from the fiducial model). Due to its robustness, our algorithm offers a new way to identify the sources of suppression in the matter power spectrum, breaking the degeneracies between baryonic feedback and new physics. Finally, we present a forecast for reconstruction of the matter power spectrum combining maps of the microwave background anisotropies from a Simons Observatory-like experiment and galaxy catalogs from the Dark Energy Spectroscopic Instrument.
\end{abstract}
\maketitle



\section{Introduction}
The standard model of cosmology agrees remarkably well with observations of the early universe, despite necessitating only around six parameters. Measurements of the primary anisotropies \citep{Aghanim2020PlanckCMB,2205.10869,2506.20707,Louis2025ACTCMB} and gravitational lensing of the cosmic microwave background (CMB)~\citep{Aghanim2020PlanckLensing,2022JCAP...09..039C,Pan2018SPTLensing,2411.06000,Madhavacheril2024ACTLensing,2024ApJ...962..112Q,2025arXiv250420038Q} give a consistent picture of the growth of structure over billions of years (see ~\cite{2025RSPTA.38340025M} for a recent review). However, some observations of the late universe paint a different picture. Apart from tensions and disagreements associated with expansion measurements (e.g. \cite{Riess2022AComprenhensive,Karim2025DESIDR2}), many measurements of cosmic shear such as those from the Dark Energy Survey \cite{Abbott2022DES} and the Hyper-Suprime Cam (HSC) \cite{Li2023HSC,Dalal2023HSC} surveys show a preference for a lower level of clustering than the CMB best-fit model (though KiDS cosmic shear now agrees better with the CMB prediction \cite{Wright2025KIDS}). The level of matter clustering is often parameterized by the root-mean-squared amplitude of fluctuations smoothed on scales of $8\;\mathrm{Mpc}/h$, denoted $\sigma_8$, or by its value rescaled by the matter density $S_8=\sigma_8(\Omega_m/0.3)^{0.5}$. This discrepancy has been coined the ``$S_8$-tension", and multiple solutions have been suggested to reconcile the conflicting measurements. 

Baryonic processes such as AGN feedback, supernova feedback, and star formation are poorly understood. These effects redistribute matter on small scales and induce systematic errors when inferring $S_8$ from weak lensing data. It has been argued that models with a higher level of baryonic feedback can reconcile conflicting measurements of $S_8$ by lowering the amplitude of the matter power spectrum on small scales and at late times~(see e.g. \cite{Amon2022Anonlinear,Dalal2023HSC,Preston2024ReconstructingThe,Perez2025ReconstructingThe,Doux2025GoingBeyond,Terasawa2025ExploringThe}).

The kinetic Sunyaev-Zeldovich (kSZ) effect is a secondary CMB anisotropy resulting from the Doppler shift of CMB photons due to Thomson scattering off free electrons moving along the line of sight. We define the electron radial momentum as
\begin{align}
    q_r(\mathbf{x}) &\equiv \left[1+\delta_e(\mathbf{x}) \right] v_r(\mathbf{x}) ,
\end{align}
where $\delta_e$ and $v_r$ denote the electron density contrast and radial peculiar velocity. The electron velocity is dominated by the large-scale linear velocity that relates to the matter density contrast ($\delta_m$) by inverting the Laplacian in Poisson's equation, giving
\begin{align}
    v_r(\mathbf{k}) = i\mu \frac{faH}{k} \delta_m(\mathbf{k}),
\end{align}
 where $k$ is the comoving wavenumber, $\mu$ is the cosine of the angle between the wavenumber and the line-of-sight vector, $f$ is the logarithmic linear growth rate, $a$ is the cosmological scale factor, and $H$ is the Hubble expansion rate. The temperature fluctuations induced by non-zero $q_r$ in the direction $\hat{\mathbf{n}}$ follow
\begin{align}
    \Delta T_\mathrm{kSZ}(\hat{\mathbf{n}}) = \int_0^{z_*} dz\; \frac{d\chi}{dz} K( z)q_r(\hat{\mathbf{n}}, z),
\end{align}
where $\chi$ is the comoving distance to the observer and
\begin{align}
    K(z) = - T_\mathrm{CMB} \bar{n}_{e, 0} \sigma_\mathrm{T} (1+z)^2 e^{-\tau (z)},
\end{align}
where $T_\mathrm{CMB}$ is the background (average) CMB temperature, $\bar{n}_{e, 0}$ is the comoving mean free electron number density at the present day, $\sigma_\mathrm{T}$ is the Thomson cross-section, and $\tau$ is the mean optical depth for electron scattering.

The kSZ effect can be used as a probe of baryonic feedback; extracted through velocity-weighted stacking of the CMB map at the location of galaxies~\citep{Tanimura2021DirectDetection,Schaan2016EvidenceFor,Raghunathan2024FirstConstraints,Schaan2021CombinedKinematic,Hadzhiyska2024EvidenceFor,Ried2024VelocityReconstruction}, it has been successfully used in combination with weak gravitational lensing~\citep{Bigwood2024WeakLensing}.  Recent studies involving multiple probes show a preference for stronger feedback models in simulation suites~(e.g. \cite{Schneider2022ConstrainingBaryonic,Preston2023ANon-linear,Dalal2025DecipheringBaryonic,Kovac2025BaryonificationII,Reischke2025AFirst,Bigwood2025TheKinetic,McCarthy2025FlamingoCombining}). When using the kSZ as a probe of feedback, the stacked electron profile is compared to the output of a diverse set of simulations. It is worth noting that the simulations used as a means for comparison are often calibrated to match the gas fraction observed in X-ray measurements~\citep{vanDaalen2011TheEffects,McCarthy2017TheBAHAMAS,Henden2018TheFABLE,vanDaalen2020ExploringThe,Salcido2023SPk,Schaye2023TheFLAMINGO}. Recent works such as \cite{Siegel2025JointXray} suggest that the X-ray gas fraction may need to be revised to account for new measurements from the eROSITA mission~\cite{Bulbul2024TheSRGeRosita}. Multiple studies report lower gas fractions in X-ray clusters than previously measured, potentially due to selection effects. Ref.~\cite{Popesso2024TheHot} cautions, however, that current simulations with enhanced feedback models also predict excessive levels of quenching for galaxies in high-mass halos. 

When using the kSZ as a probe of feedback, the selection of galaxies must match the properties of the survey to ensure a fair comparison. Thus, matching hydrodynamical simulations to kSZ stacking observations can be a challenging task. The comparison to simulations is also generally done individually for each feedback model. An exception to this, presented in \cite{Bigwood2025TheKinetic}, leverages the 400 realizations in the ANTILLES simulation suite (see \cite{Salcido2023SPk}). One limitation of this dataset is that, due to their small box size, the ANTILLES simulations cannot capture variations in the cosmic velocity field. This is because the velocities have a correlation length comparable to the simulation box size of $100\;\mathrm{Mpc}/h$. In this study, we present an alternative approach allowing for the inference of the matter power spectrum shape while interpolating between a limited number of hydrodynamical simulations with very large box sizes. Our method enables us to marginalize over the galaxy selection effect and jointly fit for the galaxy angular power spectrum with the kSZ signal (which we measure in spherical harmonics space from a template, rather than using aperture photometry).

The kSZ effect has increasingly been used as an alternate probe of feedback. The precision of kSZ measurements is set to improve significantly in the near future, with a current detection significance of $\sim10\sigma$ forecasted to reach $\sim100\sigma$ with Stage-IV galaxy surveys~\cite{Smith2018kSZTomography}. While a direct comparison of electron profiles in halos from individual simulations allows for a qualitative assessment, a model that enables more direct sampling of the shape of the power spectrum is a natural next step. Following this line of thought, our emulator framework proves it is possible to infer the shape of the matter power spectrum while accurately and continuously interpolating between a limited set of hydrodynamical simulations. This work is set to be the first of a series leading to the release of matter power spectrum shape constraints independent of gravitational lensing. Our approach will enable enhanced constraints on new physics by breaking the degeneracy between baryonic feedback and other sources of structure suppression.

\subsection{Template Approach for kSZ}

The template method for extracting the kSZ signal consists of creating a galaxy momentum map and cross-correlating it with a map of the CMB~\citep{Ho2009FindingThe,Shao2011KineticSunyaev}. Since the primary CMB and the instrument noise do not correlate with the galaxy line-of-sight velocity field, the cross-correlation process extracts the kSZ contribution from the map. Data analyses so far have performed velocity-weighted stacking of CMB maps at the location of galaxies. We demonstrate here on simulations the harmonic-space equivalent of the traditional stacking analysis. The cross-correlation bandpowers we construct $C_{\ell}$ are significantly easier to work with since their covariance matrix is nearly diagonal. (See for instance, Ref.~\citep{Wayland2025DetailedTheoretical} for more details on the modeling of the Fourier-space approach.)

To build a template, we begin by computing the galaxy density field on the sky, defined as
\begin{align}
    \delta_g(\hat{\mathbf{n}}) = \frac{n_g(\hat{\mathbf{n}})}{\bar n_g} -1,
\end{align}
where $\hat{\mathbf{n}}$ is the unit vector of an angle on the sky and $\bar n_g$ is the mean number density of galaxies over the full sky. For the velocity field, we average the radial velocities of the galaxies that fall within the same pixels. Once we have these two maps, we generate the galaxy momentum template by multiplying them on a pixel-by-pixel basis such that
\begin{align}
    \hat{b}(\hat{\mathbf{n}}) =  \delta_g(\hat{\mathbf{n}}) v_r(\hat{\mathbf{n}}).
\end{align}
The Doppler $b$ parameter captures the same information as the kSZ, and the two can be related via
\begin{align}
    b \equiv -\frac{\Delta T_\mathrm{kSZ}}{T_\mathrm{CMB}}.
\end{align}

The kSZ effect induces a bispectrum between large-scale structure and the CMB. As shown in~\cite{Smith2018kSZTomography}, this bispectrum is dominated by the squeezed limit, and we separate the signal into two relevant scales: the small-scale electron density fluctuations at $k_S \gtrsim 1$ and the large-scale cosmic velocity field at $k_L \lesssim 0.1$. The correlation between the template and the true kSZ can be calculated from the true and estimated field power spectra. Defining $\hat{q}_r(\mathbf{x}) \equiv \delta_g(\mathbf{x}) v_r(\mathbf{x})$, and following~\cite{Foreman2023SubtractingThe}, this gives 
\begin{align}
    P_{\hat q_r q_r}(k_S,  z) \approx P_{ge}(k_S, z)   \int \frac{d k_L}{4\pi^2} \int d\mu_L\; P_{v_r v_r}(k_L, \mu_L, z), 
\end{align}
where $\mu_L$ is the cosine of the angle between the large-scale mode and the line of sight. Here, we assume that the galaxy radial velocities are known. In practice, they are estimated through an initial density field reconstruction procedure. We discuss the impact of imperfections in the reconstruction in Appendix~\ref{app:systematics}. The linear radial velocity power spectrum is obtained from the linear matter power spectrum ($P_{mm}$) through
\begin{align}
    P_{v_r v_r}(k_L, \mu_L, z) =\left(i\mu_L\frac{faH}{k_L}\right)^2 P_{mm}(k_L, z).
\end{align}
The small-scale galaxy-electron cross-spectrum can be computed using a halo model approach, giving (on one-halo dominated scales)
\begin{align}
    P_{ge}(k_S, z) = \int &dM_h \; n(M_h) \frac{M_h}{\bar{\rho}_m}u_\mathrm{gas}(k_S, z; M_h) \nonumber \\ &\times \frac{1}{\bar{n}_g} \left[\langle N_\mathrm{cen}\rangle(M_h)+\langle N_\mathrm{sat}\rangle(M_h)u_\mathrm{sat}(k_S, z; M_h) \right],
\end{align}
where $\langle N_\mathrm{cen/sat}\rangle$ are the expected number of central and satellite galaxies in halos of mass $M_h$, $n$ is the comoving number density of halos of a given mass, and $\bar{\rho}_m$ is the mean comoving matter density. Also, $u_\mathrm{sat}$ is the Fourier transform of the satellite galaxies' number density as a function of scale, and $u_\mathrm{gas}$ is the Fourier transform of the gas density profile measured by the stacking approach to kSZ. The theoretical expectation value of the measured cross-spectrum is given by
\begin{align}
    C_\ell^{ \hat b b}&= \int_{z_\mathrm{min}}^{z_\mathrm{max}} \frac{dz}{\chi^2(z)} \frac{d\chi}{dz}\frac{K(z)}{T_\mathrm{CMB}} P_{\hat q_r q_r}\left(\frac{\ell+1/2}{\chi(z)}, z\right)\label{eq:limber}\\
    &= \int_{z_\mathrm{min}}^{z_\mathrm{max}} \frac{dz}{\chi^2(z)} \frac{d\chi}{dz}\frac{\sigma^2_{v_r}(z)K(z)}{T_\mathrm{CMB}} P_{ge}\left(\frac{\ell+1/2}{\chi(z)}, z\right),
\end{align}
where we define the variance in the large-scale velocity fluctuations $4\pi^2\sigma^2_{v_r} \equiv \int dk_Ld\mu_L \;P_{v_rv_r}$. The kSZ spectrum also holds information about the baryon fraction in halos through the gas profile $u_\mathrm{gas}$.  Measuring the kSZ cross-spectrum with the template method thus provides a means to determine the halo baryon fraction as a function of halo mass.

\begin{figure*}
    \centering
    \includegraphics[width=1.\linewidth]{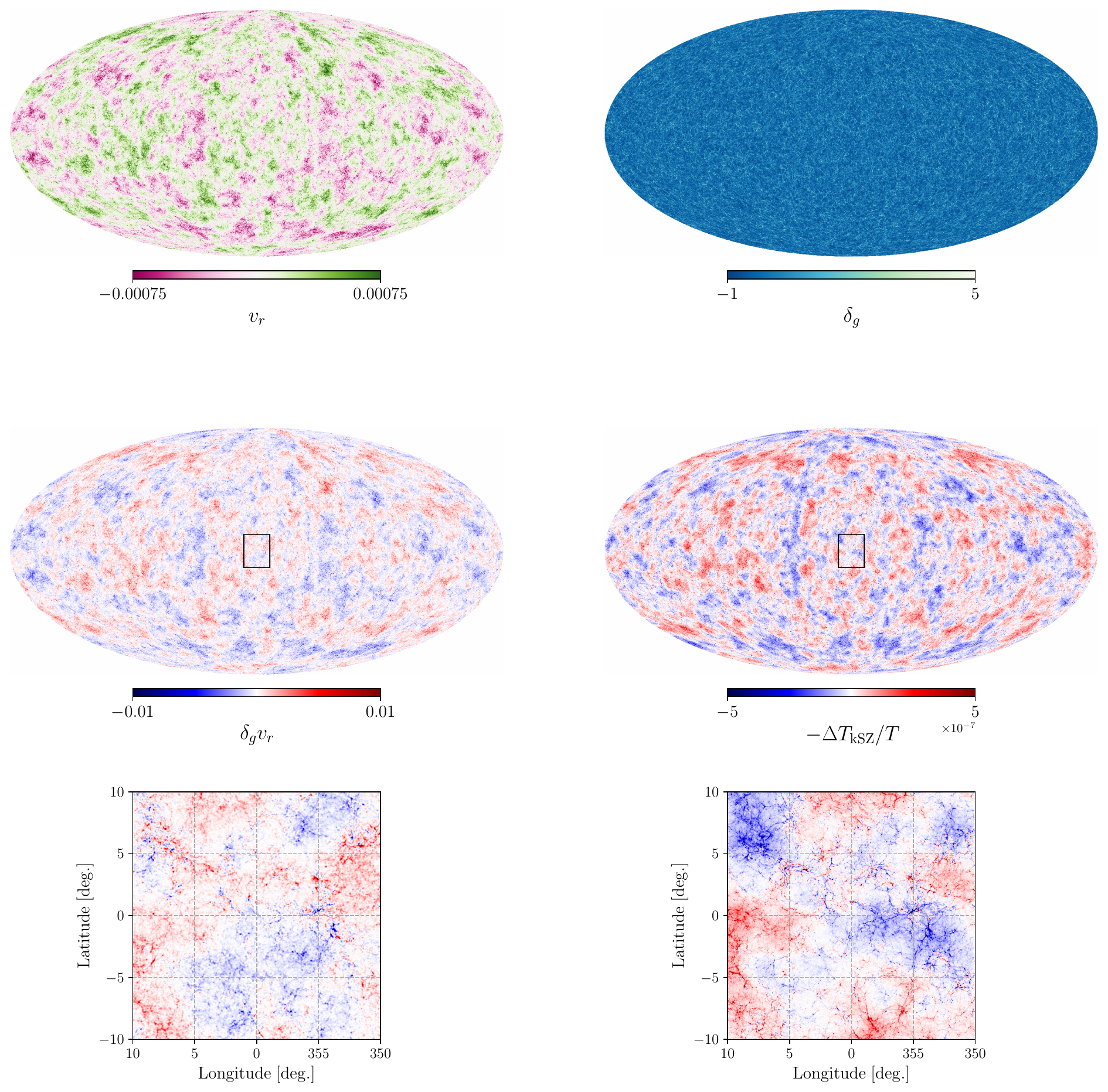}
    \caption{Example of maps used with the template method. All the maps are based on the simulation L1\textunderscore m9 at redshift $z=0.6$. By zooming-in, we can observe the correlated features of the Doppler $b$ map and the galaxy template.
    (\textit{Upper left}) Galaxy radial velocity.
    (\textit{Upper right}) Galaxy density.
    (\textit{Center left}) Doppler $b$ template.
    (\textit{Center right}) True Doppler $b$.
    (\textit{Lower left}) 20 by 20 degree region of template map.
    (\textit{Lower right}) 20 by 20 degree region of true Doppler $b$ map. The $v_r, \;\delta_g$, and $\delta_g v_r$ maps have been smoothed with a Gaussian with a FWHM of 5 arcminutes for ease of visualization.
    }
    \label{fig:six-maps}
\end{figure*}

\section{Methodology}
We develop a pipeline to parameterize and quantify the impact of baryonic feedback on the shape of the matter power spectrum as a function of the observed kSZ effect. An increasing number of emulators and fitting functions for the non-linear matter power spectrum, accounting for baryonic effects, exist in the literature. These include Gaussian processes \citep{Rogers2021GeneralFramework,Mootoovaloo2022KernelBased,Schaller2025TheFLAMINGO}, fitting functions \citep{Salcido2023SPk}, and neural networks \citep{Arico2021BACCO}. All the emulators in the literature take as input cosmological and surrogate baryonic feedback parameters (such as the baryon fraction in halos). The baryonic feedback parameters are often defined as a function of the subgrid models for the simulations on which the emulator was trained. While helpful to marginalize over feedback parameters, these models take as input values that cannot be directly measured, such as the AGN feedback gas heating or the halo baryon fraction. On the other hand, our framework is the first to infer the impact of baryonic feedback directly from cosmological observations.

\subsection{FLAMINGO Simulations}
The Virgo Consortium’s FLAMINGO suite is a set of hydrodynamical simulations with large cosmological boxes and relatively high resolution~\citep{Schaye2023TheFLAMINGO,Kugel2023FLAMINGOCalibrating}. The FLAMINGO suite encompasses a diverse range of baryonic feedback models. Notably, all the hydrodynamical simulations have dark matter-only (DMO) counterparts. This allows for the computation of the matter power spectrum suppression, which we define as
\begin{align}
    R(k) \equiv \frac{P(k)}{P_\mathrm{DMO}(k)},
\end{align}
where $P$ and $P_\mathrm{DMO}$ denote the total matter power spectra of their respective simulations. The large dataset contains maps of the kSZ effect and detailed halo catalogs with associated galaxy properties. The kSZ effect is stored on the lightcone with the Doppler $b$ parameter
\begin{align}
    b_\mathrm{sim} &= \left(\frac{n_\mathrm{e} m_\mathrm{g} \sigma_\mathrm{T}}{\Omega^2_\mathrm{pixel}d_\mathrm{A}^2\rho }\right) v_r,
\end{align}
where $m_\mathrm{e}$ is the electron mass, $n_\mathrm{e}$ is
the free electron number density, $m_\mathrm{g}$ is the gas particle mass, $\rho$ is the gas density, $\Omega_\mathrm{pixel}$ is the solid angle
of a HEALPix pixel, and $d_\mathrm{A}$ is the angular diameter distance to the observer\footnote{The units as described in \cite{Schaye2023TheFLAMINGO} include the speed of light, which we set to unity.}.
\begin{figure*}
    \centering
    \includegraphics[width=0.7\linewidth]{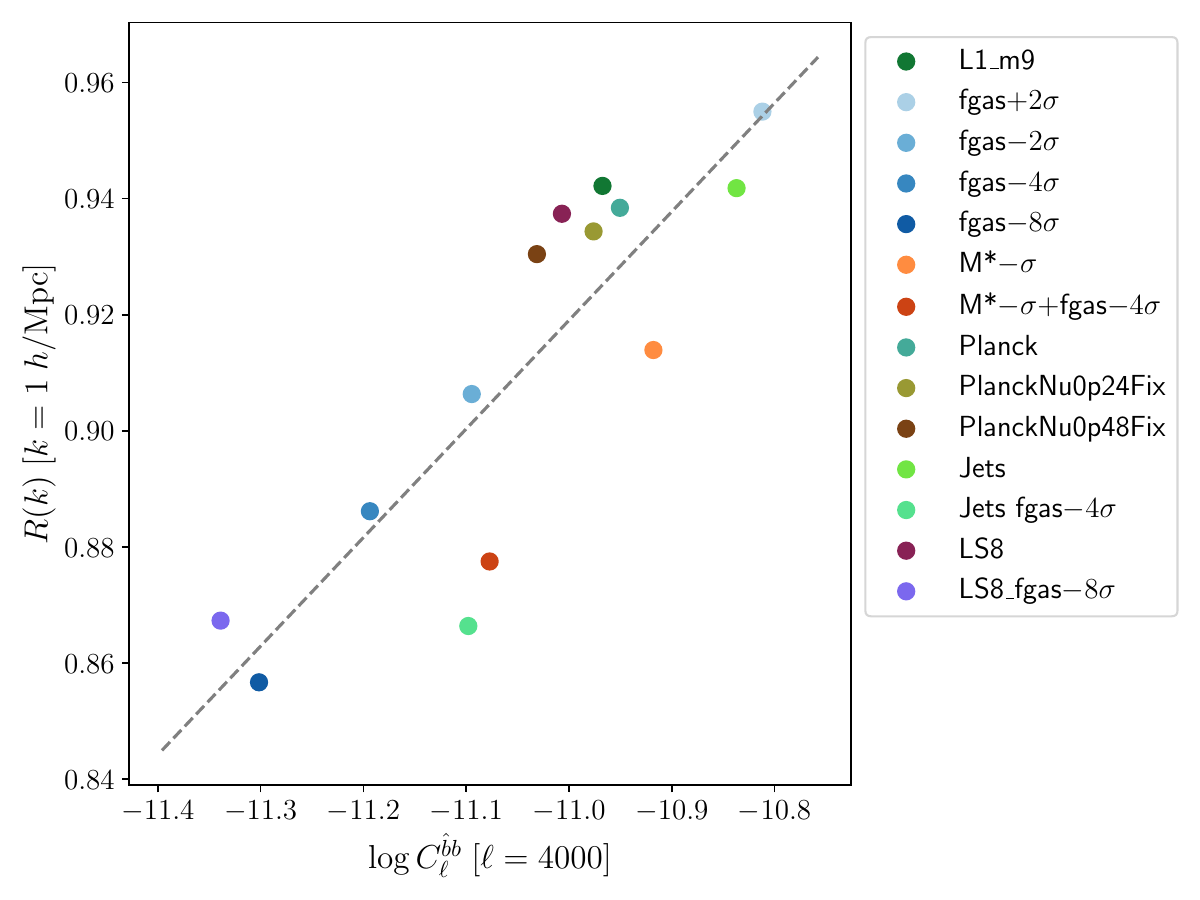}
    \caption{Variation in the matter power spectrum suppression at $k=1\;h$/Mpc as a function of the kSZ cross-correlation at $\ell=5000$ and redshift $z=0.3$. The template is formed of galaxies with stellar masses above $10^{11.3}\;\mathrm{M}_\odot$. The dashed grey line is the least-squares linear fit using the log-polynomial model.}
    \label{fig:van-daalen-plot}
\end{figure*}

To train our emulator, we use the following simulations: L1\textunderscore m9, fgas$\pm2\sigma$, fgas$- 4\sigma$, fgas$- 8\sigma$, M*$-\sigma$, M*$-\sigma+$fgas$-4\sigma$, Jet, Jet fgas$-4\sigma$, Planck, PlanckNu0p24Fix, PlanckNu0p48Fix, LS8, LS8\textunderscore fgas$-8\sigma$. This allows us to explore a wide range of feedback mechanisms, account for the impact of lowering the matter power spectrum's amplitude, and check our methods against varying neutrino masses. The $\pm N\sigma$ notation refers to simulations where the cluster gas fractions were shifted by $N$ times the error on X-ray and weak lensing data compiled by \cite{Kugel2023FLAMINGOCalibrating}. 

A scaling relation between the suppression of the matter power spectrum and baryonic feedback was found in~\cite{vanDaalen2020ExploringThe}. Building on the work of \cite{Semboloni2011QuantifyingThe, Semboloni2013Effectof}, they showed that the baryon fractions of halos of different masses (defined as the baryon mass to total mass ratio, normalized to the cosmic mean of $\Omega_b/\Omega_m$) are a direct probe of the baryonic suppression of the power spectrum as a function of scale. This relationship was shown to hold for a wide range of baryonic feedback models. This information was used to devise a fitting function for the power spectrum in \cite{Salcido2023SPk}. In Fig.~\ref{fig:van-daalen-plot}, we can see that a similar correlation exists for the kSZ cross-spectrum. For the latter, the relation is more complex as it depends on the properties of the galaxy survey used to generate the template. It also depends on the effects of feedback on galaxy formation and stellar mass. This introduces scatter in the relationship between the kSZ cross-spectrum at a given multipole and the shape of the matter power spectrum. To extract information about the power spectrum despite these sources of uncertainty, we generate many galaxy catalogs for each feedback mechanism. We then train a neural network to use the angular power spectrum of the galaxies ($C_\mathrm{\ell}^{gg}$) to remove degeneracies between the template and the true electron distribution.

\begin{figure}
    \centering
    \begin{subfigure}{\linewidth}
        \centering
        \includegraphics[width=0.875\linewidth]{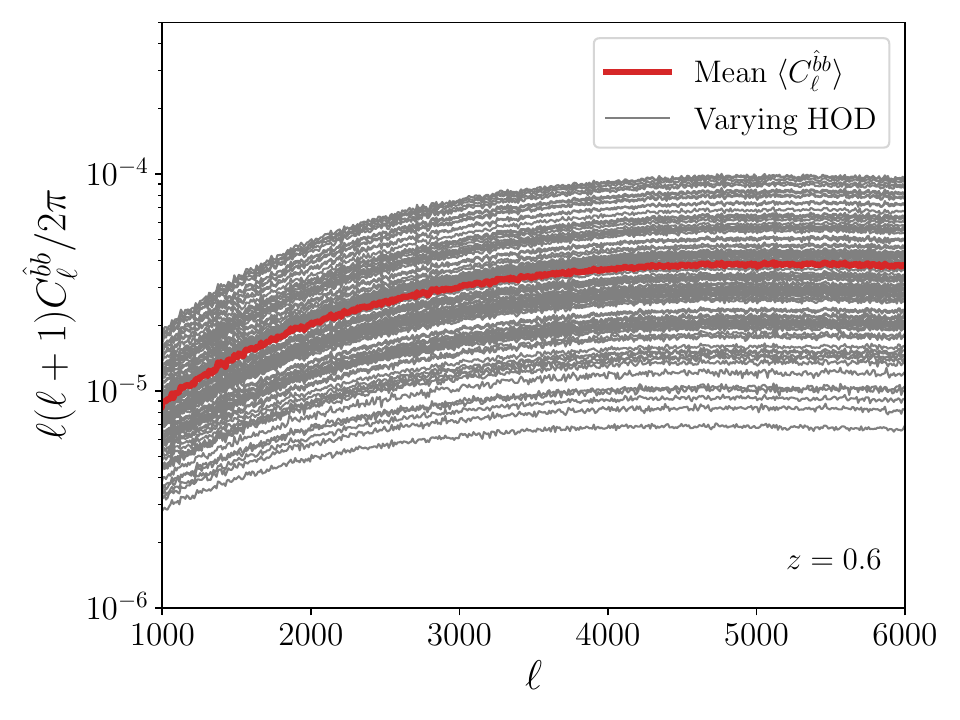}
        \caption{
        }
        \label{fig:placeholder}
    \end{subfigure}
    
    \begin{subfigure}{\linewidth}
        \centering
        \includegraphics[width=0.875\linewidth]{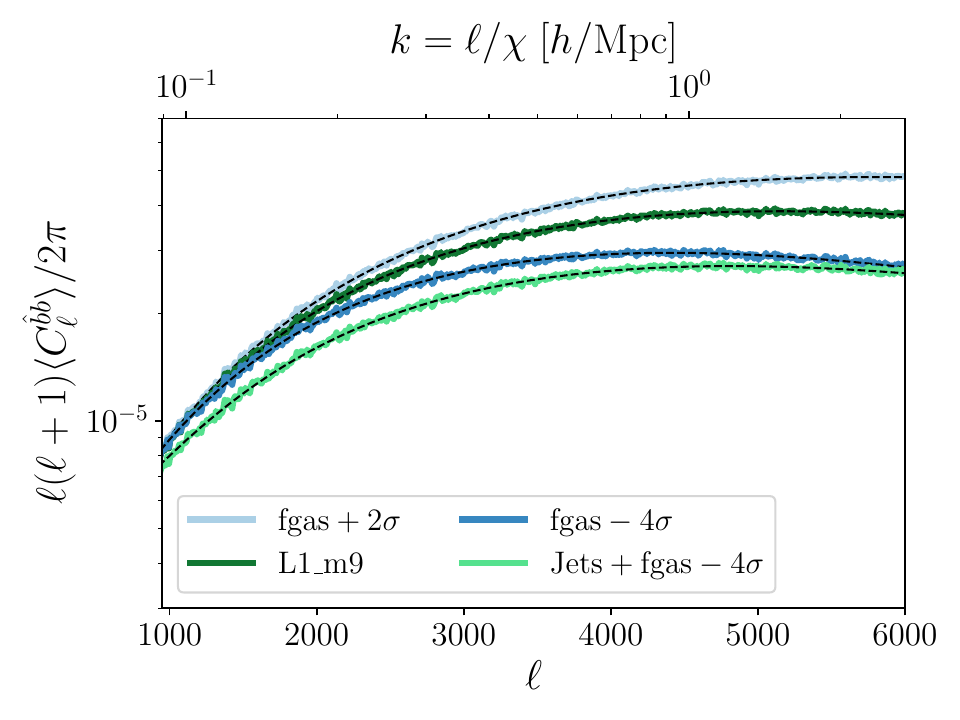}
        \caption{
        }
        \label{fig:placeholder1}
    \end{subfigure}
    
    \begin{subfigure}{\linewidth}
        \centering
        \includegraphics[width=0.875\linewidth]{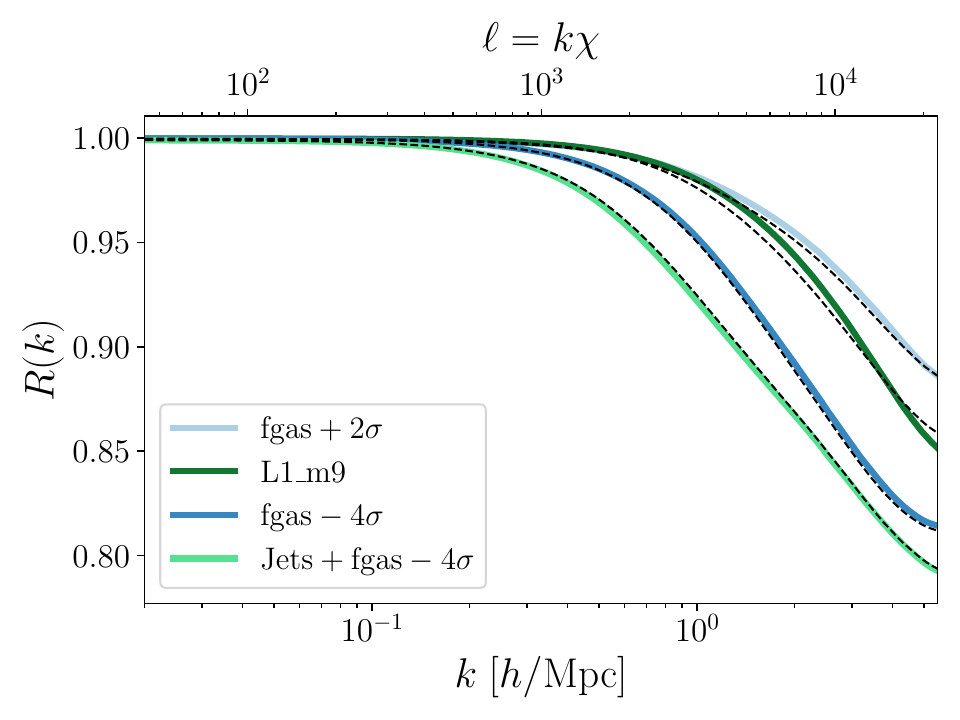}
        \caption{}
        \label{fig:placeholder2}
    \end{subfigure}
    \caption{(\textit{Top}) Doppler $b$ cross-correlation spectra from 100 HODs at redshift $z=0.6$ for the fiducial feedback model (L1$\_$m9) along with the log-polynomial fits.
    (\textit{Center}) Mean Doppler $b$ template cross-correlation power spectra. The weaker AGN feedback denoted fgas$+2\sigma$ is less suppressed than the fiducial model and the more pronounced feedback mechanisms (L1\textunderscore m9, fgas$-4\sigma$ Jets+fgas$-4\sigma$). The dashed lines represent the mean of log-polynomial fits to the cross-spectra.
    (\textit{Bottom}) Power spectrum ratio at $z=6$ for the four models shown above. The dashed lines denote the mean of the $R(k)$ predictions from our emulator. In the two bottom plots, the wavenumbers and angular multipoles are related through the Limber approximation (see Eq.~\ref{eq:limber}). All panels are at redshift $z=0.6$. \label{fig:three-panel}}
\end{figure}

\subsection{Training set and feature selection\label{sec:features}}
To generate simulated templates, we populate halos using a halo occupation distribution (HOD)~\citep{Zheng2007GalaxyEvolution} following the prescription of~\cite{Yuan2022StringentSigma8}. We calculate the probability of a halo of mass $M_h$ (defined such that the halo average density is 200 times the mean matter density) hosting a central galaxy and a number of satellites with
\begin{align}
    \langle N_\mathrm{cen}\rangle(M_h) &= \frac{f_\mathrm{ic}}{2}\mathrm{erfc}\left(\frac{\log M_\mathrm{cut}/M_h}{\sqrt{2}\sigma}\right),\\
    \langle N_\mathrm{sat}\rangle(M_h) &= \left(\frac{M_h-\kappa M_\mathrm{cut}}{M_1}\right)^\alpha \langle N_\mathrm{cen}\rangle(M_h),
\end{align}
where erfc is the complement error function and the parameters of the HOD are defined in Table~\ref{tab:hod}. For each simulation and redshift slice, we sample 100 sets of HOD parameters from a Latin Hypercube. From our set of 14 feedback models and 22 redshift slices over $0.2\leq z\leq 1.25$, this gives a total of 30800 sample points. We display the samples for the feedback model L1\textunderscore m9 at redshift $z=0.6$ in Fig.~\ref{fig:placeholder}. 
\begin{table}
    \centering
    \caption{Ranges of the HOD parameters used to generate the templates in the training and test sets. The halo masses are in units of $\mathrm{M}_\odot/h$.}
    \label{tab:hod}
    \begin{tabular}{c c c}\hline\hline
       Parameter  & Meaning  & Range \\\hline
        $\log M_\mathrm{cut}$ & Mean halo mass with a central & $[12.3, \;13.4]$ \\
        $\log M_\mathrm{1}$ & Mean halo mass with a satellite  & $[13.5,\; 14.6]$ \\
        $\log \sigma$ & Transition in the power law & $[-2.0, \; 0.0]$ \\
        $\alpha$ & Slope of the power law & $[0.5, \;1.0]$ \\
        $\kappa$ & Minimum mass scaling to host a central & $[0.001,\;3.0]$ \\
        $f_\mathrm{ic}$ & Incompleteness parameter & $[0.8,\;1.0]$\\\hline
        
    \end{tabular}
\end{table}
Each HOD yields a cross-correlation Doppler $b$ power spectrum and galaxy density auto-spectrum. We bin the spectra in bins $10\;\ell$-modes in width to reduce scatter. We then fit the binned spectra with a log-polynomial model ($\hat{C}_\ell$) following
\begin{align}
    \log \hat{C}_\ell^{X} = \sum_{n=0}^3 a^X_n \left(\log \ell\right)^n,\label{eq:poly-model}
\end{align}
where $X\in \{gg, \hat b b\}$. The best-fit values are found by minimizing the squared error between the polynomial model and the observed spectra, such that
\begin{align}
    \left[a_n^X\right]^* = \underset{a_n^X}{\mathrm{argmin}} \left\{\sum_{\ell=\ell_\mathrm{min}}^{\ell_\mathrm{max}} \frac{(2\ell+1)\Delta \ell}{2}\left|\frac{\hat{C}_\ell^{X}(a_n^X) - C_\ell^{X}}{C^X_\ell}\right|^2\right\},\label{eq:knox-fit}
\end{align}
where $\Delta \ell =10$ is the bin width. The $gg$ spectrum is fit using the modes $50\leq\ell\leq 6000$ and the $\hat bb$ spectrum is fit on smaller scales using the range $1000\leq\ell\leq 6000$. As shown in Fig.~\ref{fig:three-panel}, the log-polynomial fits agree to $\sim 1$\% for the kSZ cross-spectrum. The galaxy auto-spectrum fits with the same model agree to within $\sim 10$\%. It is worth noting that the galaxies themselves provide only little information about the shape of the power spectrum on very small scales, given its shot noise (we refer the readers to Appendix~\ref{app:info}.) 
The coefficients $a^X_n$ compress the information in the power spectra and capture changes in the baryonic feedback mechanisms. They are highly correlated, have different ranges, and confound the impact of galaxy selection with variation in baryonic feedback. This can be seen in Fig.~\ref{fig:three-panel} where a collection of 100 galaxy templates (generated from different HOD choices) gives 100 different cross-correlation spectra with \emph{the same} underlying matter power spectrum. In the lower panels of Fig.~\ref{fig:three-panel}, we compare the average cross-correlation spectra of four feedback models. We can see that the shape of the spectra correlates strongly with the shape of the matter power spectrum ratio at a similar scale (where the angular scale can be mapped to the wavenumber using the Limber approximation).

We add the redshift $z$ as a feature to the set of parameters $\{a^X_n\}$. Our input vector for each $gg$ and $\hat b b$ realization comprises nine features (two cubic fits of four parameters each, and the redshift). To summarize, for this work, a sample is an individual realization of $C_\ell^{gg}$ and $C_\ell^{\hat bb}$ generated from a given simulation, at a single redshift, and with a set of unique HOD parameters. We rescale the features to have a zero mean and unit variance, and then perform a combination of principal component analysis (PCA) and linear discriminant analysis (LDA). LDA is commonly used for classification tasks \cite{908974} because it creates features that are easier to distinguish from one another. It uses labels for each of the inputs in the training set. This combination, known as discriminant analysis of principal components, enables us to identify a subset of features that are mutually uncorrelated but individually correlated with the shape of the power spectrum.  We generate labels by assigning an integer to each power spectrum following
\begin{align}
    L_i = \mathrm{int} \left[10^5 \times\mathrm{min}_k\{R_i(k)\}\right], \label{eq:lda_label}
\end{align}
where $i$ enumerates the different samples. This labeling strategy groups together the power spectrum ratio with similar amounts of suppression. The factor of $10^5$ means the numerical value of the suppression minimum is rounded to five significant figures. Using a smaller value, such $10^3$, leads to fewer labels. After testing, we find that the LDA analysis performs better with a larger set of labels. The labels $L_i$ are used to compute the covariance matrix of the samples within each class (in our case, a class is simply one choice of feedback model). Then, the new variables produced by LDA are formed by a linear combination that maximizes the variance between samples of different classes relative to the variance within each class~\citep{Hastie2009itz}. Thus, we use LDA to create a new set of features (called the discriminant coordinates) that more easily distinguish between simulations with a high level of suppression and those with little baryonic feedback.

Following the creation of the discriminant coordinates, we compute their principal components. The principal components diagonalize the covariance matrix of the new variables and are therefore statistically independent of each other. The combination of LDA and PCA provides a mapping between the measured coefficients $a_n$ and the new features $p_n$, which are independent and more sensitive to the differences between feedback models. We retain all available features, which we label as $p^i_0,\;...,\;p^i_8$ for each of the $i^\mathrm{th}$ samples.

We also pre-process the target function, $R(k)$. The method developed by Ref.~\cite{Arico2021BACCO} works very well for our purposes. The spectra are evaluated at 200 logarithmically-spaced points in the range\footnote{The FLAMINGO matter power spectra are initially in inverse Mpc units. We transition to $h$/Mpc following this step.} $0.01-8$ Mpc${}^{-1}$ and smoothed with a Savitzky-Golay filter with a width of 31 and a polynomial order of 3. Since the curves are smooth, the sample points are highly correlated. We also use PCA decomposition on the smoothed $R(k)$ to reduce the network size and simplify the inference procedure. We find that six components are sufficient to capture the variance in the shape of the matter power spectrum between different baryonic feedback models.

\subsection{Fitting and network architecture}
\begin{figure*}
    \centering
    \includegraphics[width=0.8\linewidth]{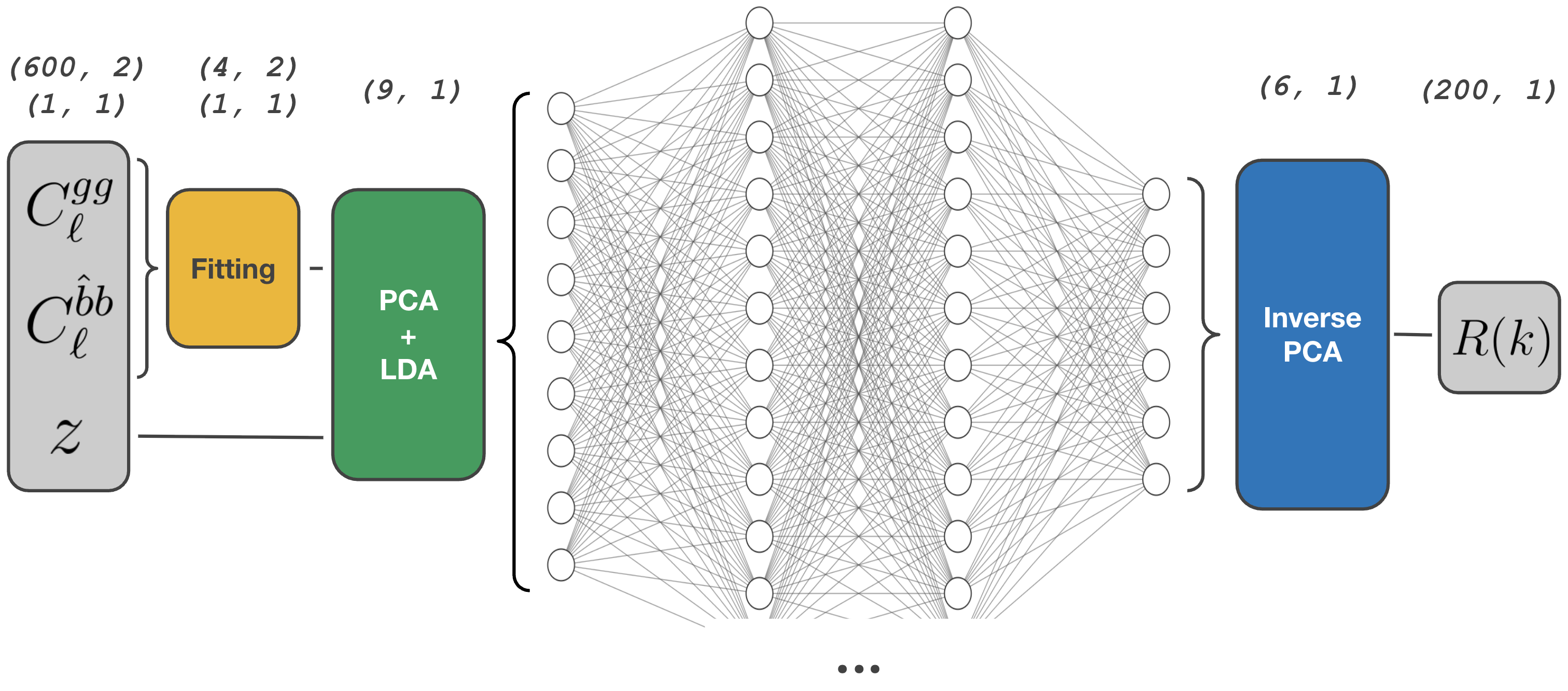}
    \caption{Workflow (from left to right) of our MLP emulator taking as input the galaxy auto-spectrum, the kSZ cross-spectrum, and the redshift. The full-size hidden layers with 1024 neurons are partially hidden for clarity. The numbers indicate the shapes of the arrays after each operation.}
    \label{fig:network}
\end{figure*}
While Fig.~\ref{fig:van-daalen-plot} illustrates the relationship between the power spectrum suppression and the kSZ amplitude, it is worth noting that this plot is only showing this correspondence for fixed stellar mass cuts, wavenumber, redshift, and angular multipole. When modeling the full matter power spectrum over a wide range of scales, with different choices of HODs and cosmic epochs, the patterns in the data become far more complex. We find that linear regression models can work well in limited cases, but do not generalize well in this context. For this reason, we opt to leverage advances in machine learning to achieve the desired model accuracy and flexibility.

For our model, we use a multi-layer perceptron (MLP). MLPs are fully connected feed-forward neural networks that can fit highly non-linear relationships. Our MLP comprises four layers: an input layer with nine neurons, two hidden layers of 1024 neurons each, and an output layer with six neurons. This configuration was found after first using linear regression methods, followed by smaller networks. We increased the number of neurons per layer to 2048 and found no improvement in performance (with slight overfitting). The workflow for the inference of the matter power spectrum suppression from the galaxy and template power spectra is outlined in Fig.~\ref{fig:network}. We use a custom loss function based on the mean absolute percent error loss function. We calculate it as
\begin{align}
    \mathcal{L}(\boldsymbol{\theta}) = \frac{1}{{N_\mathrm{train}}}\sum_{i=0}^{N_\mathrm{train}-1} \left| \frac{ \mathbf{R}^\mathrm{true}_i -\mathbf{R}^\mathrm{pred}(\boldsymbol{\theta}; p^i_0,..., p_8^i)}{\mathbf{R}^\mathrm{true}_i}\right|^2,
\end{align}
where $\mathbf{R}^\mathrm{pred}$ denotes the MLP neural network prediction and $\mathbf{R}^\mathrm{true}$ is the true vector of $R(k)$ evaluated at the sample points in $k$. Note that while the network outputs the PCA coefficients, we define the loss in terms of the full $R(k)$, as it is our target for emulation.

We train the network using the ADAM optimizer, as implemented in the \texttt{TensorFlow} library~\citep{Tensorflow}. We use a batch size of 8 and an initial learning rate of $10^{-3}$, which we reduce progressively throughout the training, lasting 60 epochs. We set a validation split of 5\%. This means that 5\% of the data is used as validation during training and changes throughout the optimization.
The validation split provides an estimate of the validation loss, which we compute after each step of the optimization process. During this optimization, we save the values of the weights and biases at the step where the validation loss is minimized. We present the validation results in the next section. 

\subsection{Model validation\label{sec:validation}}
\begin{figure}
    \centering
    \includegraphics[width=\linewidth]{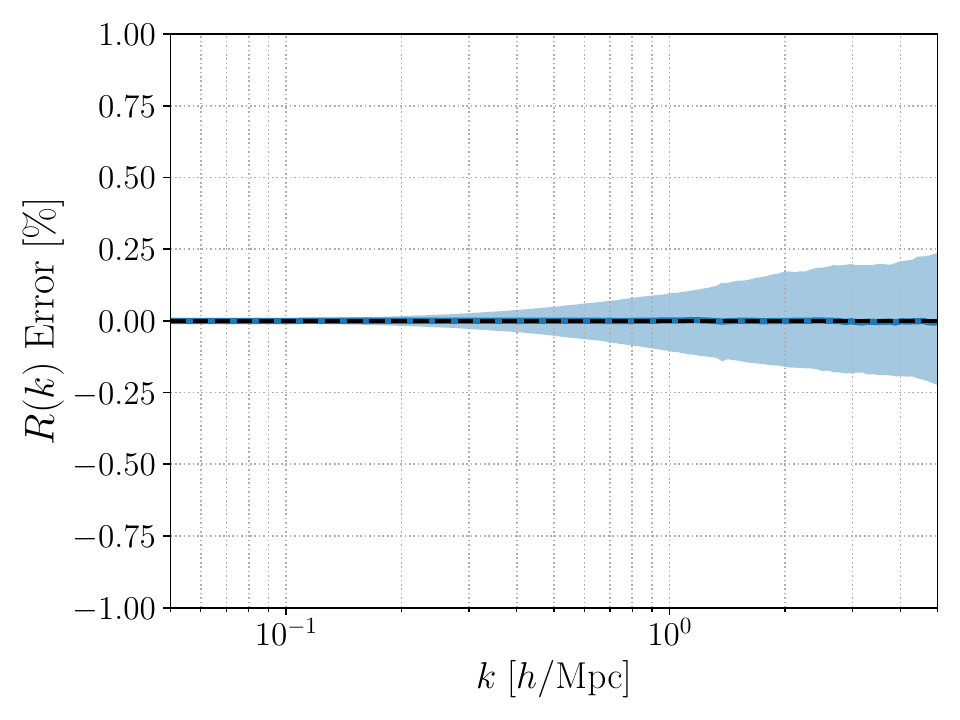}
    \caption{Test-set residuals from the neural network prediction. The test amounts for 5\% of the total dataset and includes samples from all redshift slices and feedback models. The solid line represents the median error over the validation samples while the shaded area denotes the 68\% confidence interval. The error is calculated as $R_\mathrm{model}(k)/R_\mathrm{true}(k)-1$.
    \label{fig:hod_residuals}}
    
\end{figure}

\begin{figure*}
    \centering
    \includegraphics[width=\linewidth]{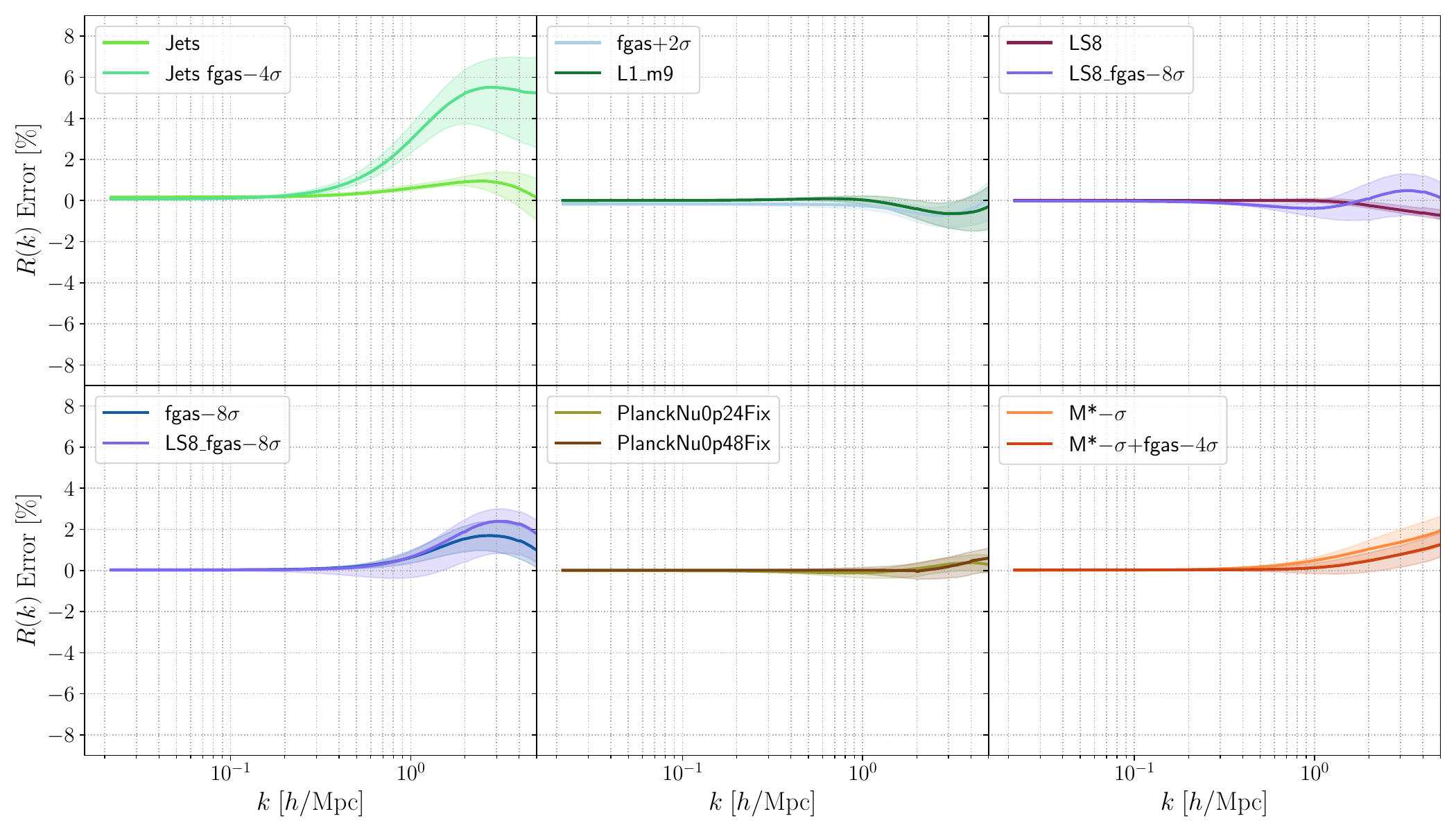}
    \caption{Set of cross-validation tests with restricted training sets. Each test is conducted by leaving two feedback models out of the training set and using the removed simulations as the test set. The simulations listed in the legend were excluded during training, and the error represents the difference between the true $R(k)$ and the network's extrapolation. The solid line represents the median error over the validation samples, while the shaded area denotes the 68\% confidence interval. The uncertainty on the matter power spectrum in the range $k\sim1-10\;h/$Mpc from CMB and weak lensing~\citep{Perez2025ReconstructingThe} is of order 15-25\%, meaning that the biases here represent $\ll 1 \sigma$ in each case.}
    \label{fig:loo_validation} 
\end{figure*}

We validate the model using three different methods. First, we split our training set, leaving 5\% for validation. 5\% of the samples are chosen at random across simulations and redshift slices. The residuals of this validation step are shown in Fig.~\ref{fig:hod_residuals}. The test set spans the full redshift range and is taken from all feedback models. We recover the matter power spectrum suppression to near percent-level accuracy over the full range of scales $k\leq 5\;h$/Mpc without bias.
Next, we conduct a cross-validation analysis. This step in the validation ensures that the emulator is not over-fitting and remains sufficiently flexible to account for diverse baryonic feedback models. The results of the set of tests are shown in Fig.~\ref{fig:loo_validation}. For these tests, we remove pairs of simulations entirely from the training set and force the network trained on the remaining simulations to extrapolate. We observe that the network maintains its accuracy even when used in a feedback mechanism outside its training set. In particular, we observe little to no bias when changing the background cosmology by lowering the $S_8$ parameter or when changing the sum of the neutrino masses. We observe some small biases when introducing variations in supernova feedback and when increasing AGN feedback to values that are very far ($8\sigma$) away from the fiducial gas-to-halo mass relation. When extrapolating to smaller amounts of feedback than what the model was trained on, we recover the power spectrum with less than 1\% bias. The single exception is the most extreme \textit{Jets} model, for which the error is close to 5\% on small scales. This model yields a power spectrum shape unlike any other in the training set since collimated jets induce a larger change in the distribution of baryons than the thermal AGN model~\citep{Schaller2025TheFLAMINGO}. This is clearly visible in Fig.~\ref{fig:van-daalen-plot}. When extrapolating to the \textit{Jets} model, while maintaining the gas-to-halo mass relation closer to the observed values, we find that our emulator extrapolates well and yields estimates that are still within 1\%. This validation test demonstrates the robustness of our algorithm to changes in feedback models and cosmology. The network emulator is remarkably stable when introducing other sources of matter power spectrum suppression, such as an increase in the sum of neutrino masses. This suggests that our approach could be used to differentiate sources of suppression that arise from changes to the linear matter power spectrum (such as warm dark matter) from late-time, non-linear effects.

\begin{figure}
    \centering
    \includegraphics[width=\linewidth]{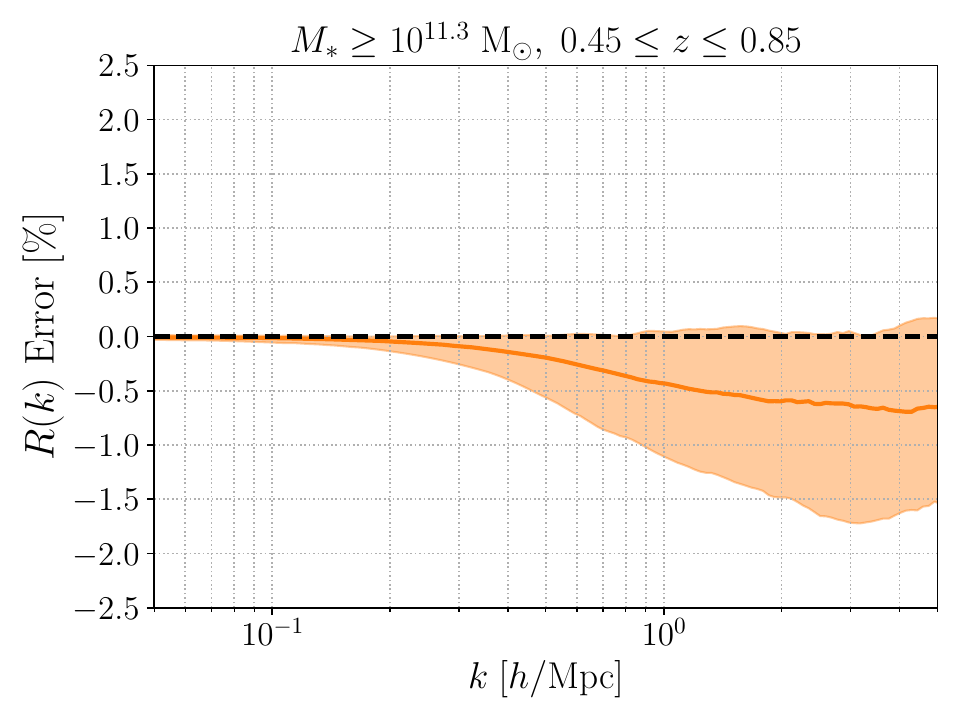}
    \caption{Validation-set residuals from the neural network prediction using a stellar mass cut rather than an HOD to create a galaxy catalog. The solid line represents the median error over the validation samples, while the shaded area denotes the 68\% confidence interval based on 308 realizations. \label{fig:stellar_residuals}}
\end{figure}

We proceed to a third round of validation by creating more realistic galaxy catalogs beyond HODs. Given that the FLAMINGO simulations are high-resolution hydrodynamical simulations, we use the stellar mass information to generate catalogs with galaxy properties correlated with feedback. Instead of generating random parameters like the HOD, we impose a completeness cut on the stellar mass of $M_*\geq10^{11.3}\;\mathrm{M}_\odot$ for samples in the redshift range $0.45<z<0.85$. This mass cut and redshift range are meant to roughly reproduce the properties of the one-percent sample of the DESI Luminous Red Galaxies, as described in \cite{Prada2025TheDESI}. We generate a realization for each feedback model and redshift slice, resulting in a total of 308 samples. The error on the prediction of our model for this alternate dataset is shown in Fig.~\ref{fig:stellar_residuals}. We observe a 0.5\% bias in the median of the matter power spectrum reconstruction, with errors reaching 3\%. This effect is less pronounced at higher redshifts and higher stellar masses. We thus attribute the increased error to the correlation between the stellar mass and feedback strength, which is not present when selecting halos based on total mass. Given the HOD parameterization (see Table~\ref{tab:hod}), the emulator training set does not account for this correlation.

The galaxy auto-spectrum and the kSZ cross-spectrum from the stellar mass cuts have a different shape than the outputs of the HOD procedure. The predictive power of the network degrades outside its training set. Nevertheless, the network still performs at the percent-level at $k\leq 0.8\;h/$Mpc and at the 2\%-level at $k\leq 5\;h/$Mpc. Crucially, the network used for this test has only been exposed to HOD-generated data in its training set. The choice of $k\leq 5\;h/$Mpc is motivated by the scale-cut imposed by the Dark Energy Survey~\citep{Doux2022DES} using the \texttt{HMCode2020} calculation for the non-linear power spectrum~\citep{Mead2021HMCODE}. Our emulator matches the accuracy of \texttt{HMCode2020} ($\approx$ 5\% on this range of scales. This is true even when extrapolating outside its training set, except for the feedback models with fgas$\leq -4\sigma$. This motivates the need to expand the training set to include other subgrid models, allowing for enhanced baryonic feedback. We leave the treatment of other simulated datasets such as ANTILLES~\citep{Salcido2023SPk}, BAHAMAS~\citep{McCarthy2017TheBAHAMAS}, FABLE~\citep{Henden2018FABLE}, Millenium TNG~\citep{Pakmor2023MilleniumTNG}, and SIMBA~\citep{Dave2019SIMBA} to future work.

\section{Results}

To illustrate the potential uses of our algorithm, we create a mock observation of $C_\ell^{gg}$ and $C_\ell^{\hat b b}$. We take a typical HOD from our validation set at redshift $z=0.6$ with the fiducial model (L1\textunderscore m9) as our data points. Instead of fitting the spectra using the cosmic variance limit errors, as we did in Eq.~\eqref{eq:knox-fit}, we compute realistic error bars accounting for partial sky coverage and uncertainty in the kSZ measurement.

\begin{figure}
    \centering
    \includegraphics[width=\linewidth]{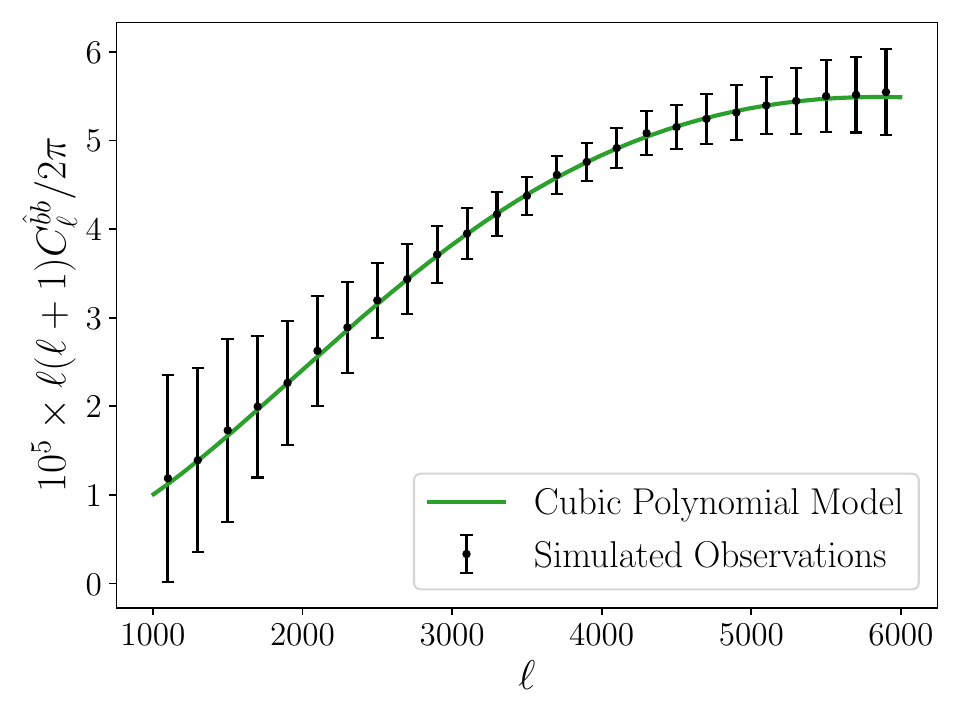}
    \caption{Simulated observations of a kSZ cross-spectrum with error bars from the primary CMB and its noise. The green line denotes the log-space cubic model of Eq.~\eqref{eq:poly-model} used to compress the data. The detection significance of the kSZ signal for this simulated measurement is $\approx50\sigma$.
    }
    \label{fig:cross_errors}
\end{figure}

To estimate the covariance matrix of the cross-spectrum, we create 750 Gaussian CMB realizations. For the primary CMB, we use the Boltzmann code CAMB~\citep{Lewis2011CAMB} and set the parameters to fit the fiducial simulation's cosmological parameters. We generate noise power spectra that are similar to those for a Simons Observatory-like experiment~\citep{Ade2019TheSimons}\footnote{This is a rough choice of noise level; we do not attempt to be precise and capture sources of noise such as the atmosphere and foreground residuals.}
\begin{align}
    N_\ell^\mathrm{CMB} = s^2_w \exp\left[\frac{\ell(\ell+1)\theta_\mathrm{FWHM}^2}{8\ln 2}\right]
\end{align}
where the beam is $\theta_\mathrm{FWHM}= 1.5$ arcminutes and the white noise level is set to $s_w=5.0\;\mu$K-arcmin. We combine the map of the primary CMB and its noise with the Doppler $b$ map (adjusting for units) to create $N_\mathrm{sims}=750$ mock sky observations including CMB, noise, and kSZ. Note that we use the same underlying Doppler $b$ for all realizations. As before, we cross-correlate the total map with the Doppler $b$ template. This provides a set of angular cross-spectra from which we estimate the errors on the $\hat {b} b$ measurement. The covariance matrix we obtain from this set has negligible off-diagonal terms, and its diagonal entries read
\begin{align}
   \sigma^2(C_\ell^{\hat b b}) = \frac{1}{N_z f_\mathrm{sky}}\frac{1}{N_\mathrm{sims}-1} \sum_i \left(C_\ell^{\hat b b+\mathrm{noise}, \;i} - \langle C_\ell^{\hat b b+\mathrm{noise}}\rangle\right)^2, \label{eq:sim-errors} 
\end{align}
where $C_\ell^{\hat b b+\mathrm{noise}}$ is the cross-spectrum of the template with the map including the non-kSZ contributions, $f_\mathrm{sky}$ is the joint fraction of the sky covered by the galaxy and CMB surveys. We take $f_\mathrm{sky}=0.2$ to match the DESI footprint overlap with SO, assuming an overlap of 10000 deg$^{2}$, following \cite{Abitbol2025SO}. This matches the redshift overlap of the first two redshift bins of the DESI LRG samples. Given the high noise level, we re-bin the cross-spectra to use 20 linearly-spaced bins in the range $\ell\in[1000, \;6000]$. The cross-spectrum used in our forecast and its error bars are shown in Fig.~\ref{fig:cross_errors}. For the galaxy auto-spectrum, we find that the relative uncertainty on $C_\ell^{gg}$ is orders of magnitude smaller than the uncertainty on $C_\ell^{\hat bb}$. Because of this limited contribution, we fix the galaxy spectrum parameters $a_n^{gg}$ in our forecast. We do the same with the redshift of the survey and vary only $a_n^{\hat bb}$.  

\begin{figure}
    \centering
    \includegraphics[width=\linewidth]{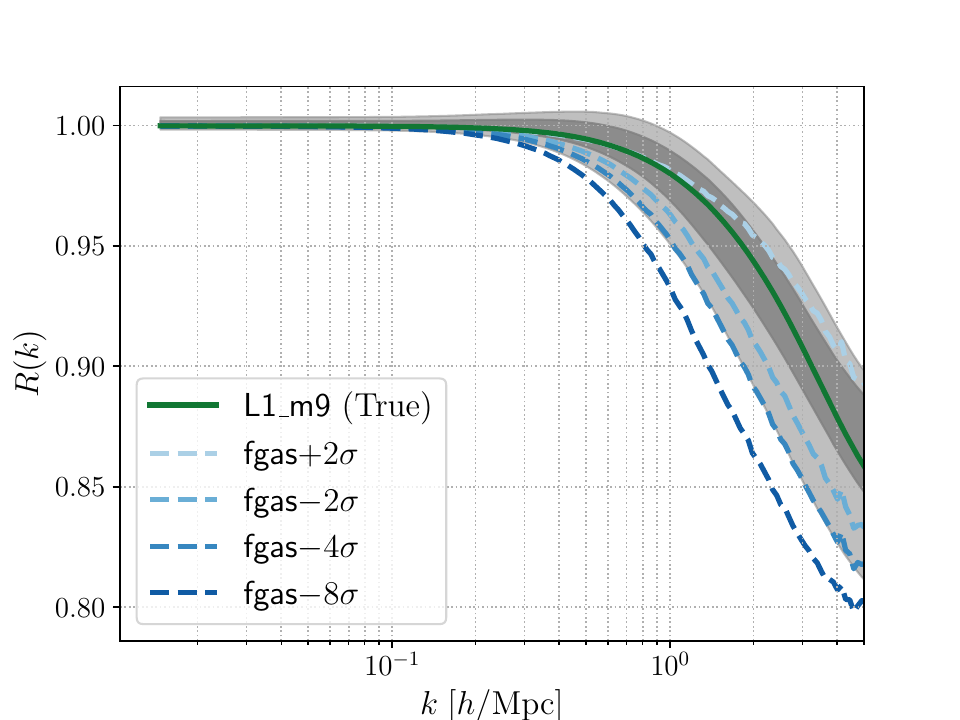}
    \caption{Inferred matter power spectrum ratio from the simulated observations of Fig.~\ref{fig:cross_errors} at $z=0.6$. The light (dark) shaded area corresponds to the 68\% (95\%) confidence interval propagated from the error margins in the kSZ cross-spectrum using MCMC sampling. This represents measurement error and should not be confused with the emulator error shown previously. The dashed lines represent the true suppression for different feedback models.
    }
    \label{fig:inference}
\end{figure}

From the error bars calculated in Eq.~\eqref{eq:sim-errors}, we construct a Gaussian likelihood for the parameters $a^{\hat bb}_n$ defined in Eq.~\eqref{eq:poly-model} which reads
\begin{align}
    \ln L (a_n^{\hat bb}) \propto -\frac{1}{2}\sum_\ell \frac{\left[\hat{C}_\ell^{\hat bb}(a_n^{\hat bb}) - C_\ell^{\hat bb}\right]^2}{\sigma^2(C_\ell^{\hat bb})}, \label{eq:gauss-like}
\end{align}
We use Markov chain Monte Carlo (MCMC) sampling to compute the distribution of the parameters given the simulated observations of $C_\ell^{\hat bb}$. We use the \texttt{emcee} package~\citep{Foreman-Mackey2013Emcee} to sample the likelihood, employing eight walkers to generate 200,000 samples each (with a burn-in of 40,000). We test convergence by ensuring that the chains are over 50 autocorrelation lengths. To avoid unphysical models, we impose a uniform prior on the factors $p_n$ and reject the samples outside a given range. The upper and lower limits are found by the bounds on these parameters in the training set. Since the $p_n$ values are independent, the priors for each parameter can be applied individually (without a need to account for correlations as would be the case if we used the prior on the $a_n$). We perform the transformations for each sample listed in Sec.~\ref{sec:features}. We then feed the remapped parameters into our emulator to obtain samples of the matter power spectrum suppression. This step is remarkably fast, as our emulator processes over $10^6$ samples in under 30 seconds during an interactive Python session. In Fig. 8, we show the 68\% confidence interval and mean of the samples. The signal to noise ratio of this mock measurement can be calculated using $\mathrm{SNR} = \sqrt{\chi^2_\mathrm{bf}-\chi^2_\mathrm{null}}$, where $-\frac{1}{2}\chi^2$ is the quantity on the right hand side of Eq.~\eqref{eq:gauss-like}, $\chi^2_\mathrm{bf}$ denotes the best-fitting model, and $\chi^2_\mathrm{null}$ is the null hypothesis with kSZ signal being zero on all scales. The signal-to-noise for the data points in Fig.~\ref{fig:cross_errors} is $\sim 50\sigma$.
This matches closely the predictions of~\cite{Smith2018kSZTomography} when considering measurements over the redshift interval $0.4\leq z\leq 0.8$. In Fig.~\ref{fig:inference}, we also display the posterior distribution on the matter power spectrum suppression. We can see that the kSZ signal allows us to discriminate between the fiducial model and the stronger feedback models (fgas$-4\sigma$, fgas$-8\sigma$) that fall outside of the 68\% confidence interval.

\section{Discussion}
We present the first power spectrum suppression emulator based on kSZ observables. We utilize the diverse set of simulations from the FLAMINGO suite and create over 30,000 mock galaxy clustering and template kSZ measurements. We generate the templates by populating dark matter halos following a halo occupation distribution method. We consider 14 feedback models in the redshift range $0.2\leq z\leq 1.25$ in bins of width $\Delta z=0.05$. We then train a multi-layer perceptron to predict the scale-dependent suppression of the matter power spectrum from baryonic feedback from these simulated datasets using a custom loss function. What sets this work apart from the emulators in the literature is that its input is directly observable. Following a similar data-driven approach, an MLP algorithm has recently been developed to constrain the epoch of reionization directly from kSZ using maps rather than angular power spectra~\citep{Farhadi2025MachineLearning}.

We develop a new approach to condense the information in the kSZ cross-correlation and galaxy number density angular power spectra through a combination of linear discriminant analysis and principal component analysis. We validate the model thoroughly in three steps. First, we remove pairs of similar feedback models from the training set and evaluate the model's performance when extrapolating to unseen data. Our model retains its accuracy for the vast majority of feedback implementations, except for the most extreme feedback simulations. Second, we generate catalogs based on a halo stellar mass threshold and evaluate the model when faced with galaxy catalogs not generated by an HOD. Finally, we ensure the model extracts the information mainly from the template kSZ and not from the redshift or galaxy density angular power spectrum. We confirm that the best predictor of the suppression of the matter power spectrum is the shape of the kSZ template angular power spectrum.

Ongoing and future surveys, such as DESI~\citep{Karim2025DESIDR2}, Euclid~\citep{Euclid2022Prep}, Roman~\citep{Spergel2015WideField}, SPHEREx~\citep{Dore2014CosmologyWith} and the Rubin Observatory's LSST~\cite{Guy2025LSST}, will produce high-resolution templates for extracting the kSZ signal. When combined with low-noise observations of the CMB by experiments such as Simons Observatory~\citep{Abitbol2025SO} and SPT-3G~\cite{Balkenhol2023SPTCMB}, we expect the total signal-to-noise ratio to exceed $100\sigma$, a factor of 10 improvement over existing studies. Our neural network allows for high-speed computations of the baryonic corrections to the matter power spectrum (with millisecond runtime). It could easily be integrated into gravitational lensing pipelines for joint fitting of the kSZ cross-spectrum and galaxy lensing. It could also be extended beyond the redshift range $ 0.2 \leq z \leq 1.25$ considered here, using the same training procedure as described above. Currently, kSZ information is included through the electron density profile in a discrete set of simulations. While this establishes an informative baseline, our approach allows for interpolation between feedback models and direct MCMC sampling. This work presents the first step in jointly fitting the kSZ effect and the clustering of galaxies and matter. It opens a new observational window that will allow us to isolate the causes of the statistical anomalies in weak gravitational lensing. It will help identify valid baryonic feedback models and potentially isolate hidden signatures of exotic dark matter candidates. While it generalizes very well to many cosmologies beyond its training set, some validation tests show that this emulator approach would benefit from training on a more diverse set of subgrid models. In future work, we will calibrate with a broader set of hydrodynamical simulations and release the inference framework for use by the community.

\section*{Acknowledgments}
MM acknowledges support from NSF grants AST-2307727 and AST-2153201. MM and AL acknowledge support from NASA grant 21-ATP21-0145.
KMS was supported by an NSERC Discovery Grant, by the Daniel Family Foundation, and by the Centre for the Universe at Perimeter Institute. Research at Perimeter Institute is supported by the Government of Canada through Industry Canada and by the Province of Ontario through the Ministry of Research \& Innovation. XC acknowledges support from the U.S.\ Department of Energy. This work used the DiRAC@Durham facility managed by the Institute for
Computational Cosmology on behalf of the STFC DiRAC HPC Facility
(\url{www.dirac.ac.uk}). The equipment was funded by BEIS capital funding via
STFC capital grants ST/K00042X/1, ST/P002293/1, ST/R002371/1 and ST/S002502/1,
Durham University and STFC operations grant ST/R000832/1. DiRAC is part of the
National e-Infrastructure.

\section*{Data Availability}
The angular power spectra of the galaxies and templates generated through this work are available upon request.


\bibliographystyle{apsrev.bst}
\bibliography{example}

@string{june = {June}}

@article{2022JCAP...09..039C,
 adsurl = {https://ui.adsabs.harvard.edu/abs/2022JCAP...09..039C},
 archiveprefix = {arXiv},
 author = {{Carron}, Julien and {Mirmelstein}, Mark and {Lewis}, Antony},
 doi = {10.1088/1475-7516/2022/09/039},
 eid = {039},
 eprint = {2206.07773},
 journal = {\jcap},
 month = {September},
 number = {9},
 pages = {039},
 primaryclass = {astro-ph.CO},
 title = {{CMB lensing from Planck PR4 maps}},
 volume = {2022},
 year = {2022}
}

@article{2024ApJ...962..112Q,
 adsurl = {https://ui.adsabs.harvard.edu/abs/2024ApJ...962..112Q},
 archiveprefix = {arXiv},
 author = {{Qu}, Frank J. and {Sherwin}, Blake D. and {Madhavacheril}, Mathew S. and {Han}, Dongwon and {Crowley}, Kevin T. and {Abril-Cabezas}, Irene and {Ade}, Peter A.~R. and {Aiola}, Simone and {Alford}, Tommy and {Amiri}, Mandana and {Amodeo}, Stefania and {An}, Rui and {Atkins}, Zachary and {Austermann}, Jason E. and {Battaglia}, Nicholas and {Battistelli}, Elia Stefano and {Beall}, James A. and {Bean}, Rachel and {Beringue}, Benjamin and {Bhandarkar}, Tanay and {Biermann}, Emily and {Bolliet}, Boris and {Bond}, J. Richard and {Cai}, Hongbo and {Calabrese}, Erminia and {Calafut}, Victoria and {Capalbo}, Valentina and {Carrero}, Felipe and {Carron}, Julien and {Challinor}, Anthony and {Chesmore}, Grace E. and {Cho}, Hsiao-mei and {Choi}, Steve K. and {Clark}, Susan E. and {C{\'o}rdova Rosado}, Rodrigo and {Cothard}, Nicholas F. and {Coughlin}, Kevin and {Coulton}, William and {Dalal}, Roohi and {Darwish}, Omar and {Devlin}, Mark J. and {Dicker}, Simon and {Doze}, Peter and {Duell}, Cody J. and {Duff}, Shannon M. and {Duivenvoorden}, Adriaan J. and {Dunkley}, Jo and {D{\"u}nner}, Rolando and {Fanfani}, Valentina and {Fankhanel}, Max and {Farren}, Gerrit and {Ferraro}, Simone and {Freundt}, Rodrigo and {Fuzia}, Brittany and {Gallardo}, Patricio A. and {Garrido}, Xavier and {Gluscevic}, Vera and {Golec}, Joseph E. and {Guan}, Yilun and {Halpern}, Mark and {Harrison}, Ian and {Hasselfield}, Matthew and {Healy}, Erin and {Henderson}, Shawn and {Hensley}, Brandon and {Herv{\'\i}as-Caimapo}, Carlos and {Hill}, J. Colin and {Hilton}, Gene C. and {Hilton}, Matt and {Hincks}, Adam D. and {Hlo{\v{z}}ek}, Ren{\'e}e and {Ho}, Shuay-Pwu Patty and {Huber}, Zachary B. and {Hubmayr}, Johannes and {Huffenberger}, Kevin M. and {Hughes}, John P. and {Irwin}, Kent and {Isopi}, Giovanni and {Jense}, Hidde T. and {Keller}, Ben and {Kim}, Joshua and {Knowles}, Kenda and {Koopman}, Brian J. and {Kosowsky}, Arthur and {Kramer}, Darby and {Kusiak}, Aleksandra and {La Posta}, Adrien and {Lague}, Alex and {Lakey}, Victoria and {Lee}, Eunseong and {Li}, Zack and {Li}, Yaqiong and {Limon}, Michele and {Lokken}, Martine and {Louis}, Thibaut and {Lungu}, Marius and {MacCrann}, Niall and {MacInnis}, Amanda and {Maldonado}, Diego and {Maldonado}, Felipe and {Mallaby-Kay}, Maya and {Marques}, Gabriela A. and {McMahon}, Jeff and {Mehta}, Yogesh and {Menanteau}, Felipe and {Moodley}, Kavilan and {Morris}, Thomas W. and {Mroczkowski}, Tony and {Naess}, Sigurd and {Namikawa}, Toshiya and {Nati}, Federico and {Newburgh}, Laura and {Nicola}, Andrina and {Niemack}, Michael D. and {Nolta}, Michael R. and {Orlowski-Scherer}, John and {Page}, Lyman A. and {Pandey}, Shivam and {Partridge}, Bruce and {Prince}, Heather and {Puddu}, Roberto and {Radiconi}, Federico and {Robertson}, Naomi and {Rojas}, Felipe and {Sakuma}, Tai and {Salatino}, Maria and {Schaan}, Emmanuel and {Schmitt}, Benjamin L. and {Sehgal}, Neelima and {Shaikh}, Shabbir and {Sierra}, Carlos and {Sievers}, Jon and {Sif{\'o}n}, Crist{\'o}bal and {Simon}, Sara and {Sonka}, Rita and {Spergel}, David N. and {Staggs}, Suzanne T. and {Storer}, Emilie and {Switzer}, Eric R. and {Tampier}, Niklas and {Thornton}, Robert and {Trac}, Hy and {Treu}, Jesse and {Tucker}, Carole and {Ullom}, Joel and {Vale}, Leila R. and {Van Engelen}, Alexander and {Van Lanen}, Jeff and {van Marrewijk}, Joshiwa and {Vargas}, Cristian and {Vavagiakis}, Eve M. and {Wagoner}, Kasey and {Wang}, Yuhan and {Wenzl}, Lukas and {Wollack}, Edward J. and {Xu}, Zhilei and {Zago}, Fernando and {Zheng}, Kaiwen},
 doi = {10.3847/1538-4357/acfe06},
 eid = {112},
 eprint = {2304.05202},
 journal = {\apj},
 month = {February},
 number = {2},
 pages = {112},
 primaryclass = {astro-ph.CO},
 title = {{The Atacama Cosmology Telescope: A Measurement of the DR6 CMB Lensing Power Spectrum and Its Implications for Structure Growth}},
 volume = {962},
 year = {2024}
}

@article{2025arXiv250420038Q,
 adsurl = {https://ui.adsabs.harvard.edu/abs/2025arXiv250420038Q},
 archiveprefix = {arXiv},
 author = {{Qu}, Frank J. and {Ge}, Fei and {Kimmy Wu}, W.~L. and {Abril-Cabezas}, Irene and {Madhavacheril}, Mathew S. and {Millea}, Marius and {Anderes}, Ethan and {Anderson}, Adam J. and {Ansarinejad}, Behzad and {Archipley}, Melanie and {Atkins}, Zachary and {Balkenhol}, Lennart and {Battaglia}, Nicholas and {Benabed}, Karim and {Bender}, Amy N. and {Benson}, Bradford A. and {Bianchini}, Federico and {Bleem}, Lindsey. E. and {Bolliet}, Boris and {Bond}, J Richard and {Bouchet}, Fran{\c{c}}ois. R. and {Bryant}, Lincoln and {Calabrese}, Erminia and {Camphuis}, Etienne and {Carlstrom}, John E. and {Carron}, Julien and {Challinor}, Anthony and {Chang}, Clarence L. and {Chaubal}, Prakrut and {Chen}, Geoff and {Chichura}, Paul M. and {Choi}, Steve K. and {Chokshi}, Aman and {Chou}, Ti-Lin and {Coerver}, Anna and {Coulton}, William and {Crawford}, Thomas M. and {Daley}, Cail and {Darwish}, Omar and {de Haan}, Tijmen and {Devlin}, Mark J. and {Dibert}, Karia R. and {Dobbs}, Matthew A. and {Doohan}, Michael and {Doussot}, Aristide and {Duivenvoorden}, Adriaan J. and {Dunkley}, Jo and {Dunner}, Rolando and {Dutcher}, Daniel and {Embil Villagra}, Carmen and {Everett}, Wendy and {Farren}, Gerrit S. and {Feng}, Chang and {Ferraro}, Simone and {Ferguson}, Kyle R. and {Fichman}, Kyra and {Finson}, Emily and {Foster}, Allen and {Gallardo}, Patricio A. and {Galli}, Silvia and {Gambrel}, Anne E. and {Gardner}, Rob W. and {Goeckner-Wald}, Neil and {Gualtieri}, Riccardo and {Guidi}, Federica and {Guns}, Sam and {Halpern}, Mark and {Halverson}, Nils W. and {Hill}, J. Colin and {Hilton}, Matt and {Hivon}, Eric and {Holder}, Gilbert P. and {Holzapfel}, William L. and {Hood}, John C. and {Howe}, Doug and {Hryciuk}, Alec and {Huang}, Nicholas and {Hubmayr}, Johannes and {K{\'e}ruzor{\'e}}, Florian and {Khalife}, Ali R. and {Kim}, Joshua and {Knox}, Lloyd and {Korman}, Milo and {Kornoelje}, Kayla and {Kosowsky}, Arthur and {Kuo}, Chao-Lin and {Jense}, Hidde T. and {La Posta}, Adrien and {Levy}, Kevin and {Lowitz}, Amy E. and {Louis}, Thibaut and {Lu}, Chunyu and {Lynch}, Gabriel P. and {MacCrann}, Niall and {Maniyar}, Abhishek and {Martsen}, Emily S. and {McMahon}, Jeff and {Menanteau}, Felipe and {Montgomery}, Joshua and {Nakato}, Yuka and {Moodley}, Kavilan and {Namikawa}, Toshiya and {Natoli}, Tyler and {Niemack}, Michael D. and {Noble}, Gavin I. and {Omori}, Yuuki and {Ouellette}, Aaron and {Page}, Lyman A. and {Pan}, Zhaodi and {Paschos}, Pascal and {Phadke}, Kedar A. and {Pollak}, Alexander W. and {Prabhu}, Karthik and {Quan}, Wei and {Raghunathan}, Srinivasan and {Rahimi}, Mahsa and {Rahlin}, Alexandra and {Reichardt}, Christian L. and {Riebel}, Dave and {Rouble}, Maclean and {Ruhl}, John E. and {Schaan}, Emmanuel and {Schiappucci}, Eduardo and {Sehgal}, Neelima and {Sierra}, Carlos E. and {Simpson}, Aidan and {Sherwin}, Blake D. and {Sif{\'o}n}, Crist{\'o}bal and {Spergel}, David N. and {Staggs}, Suzanne T. and {Sobrin}, Joshua A. and {Stark}, Antony A. and {Stephen}, Judith and {Tandoi}, Chris and {Thorne}, Ben and {Trendafilova}, Cynthia and {Umilta}, Caterina and {Van Engelen}, Alexander and {Vieira}, Joaquin D. and {Vitrier}, Aline and {Wan}, Yujie and {Whitehorn}, Nathan and {Wollack}, Edward J. and {Young}, Matthew R. and {Zebrowski}, Jessica A.},
 doi = {10.48550/arXiv.2504.20038},
 eid = {arXiv:2504.20038},
 eprint = {2504.20038},
 journal = {arXiv e-prints},
 month = {April},
 pages = {arXiv:2504.20038},
 primaryclass = {astro-ph.CO},
 title = {{Unified and consistent structure growth measurements from joint ACT, SPT and \textbackslashtextit\{Planck\} CMB lensing}},
 year = {2025}
}

@article{2025RSPTA.38340025M,
 adsurl = {https://ui.adsabs.harvard.edu/abs/2025RSPTA.38340025M},
 archiveprefix = {arXiv},
 author = {{Madhavacheril}, Mathew S.},
 doi = {10.1098/rsta.2024.0025},
 eid = {20240025},
 eprint = {2411.08152},
 journal = {Philosophical Transactions of the Royal Society of London Series A},
 month = {February},
 number = {2290},
 pages = {20240025},
 primaryclass = {astro-ph.CO},
 title = {{Assessing the growth of structure over cosmic time with cosmic microwave background lensing}},
 volume = {383},
 year = {2025}
}

@article{2205.10869,
 adsurl = {https://ui.adsabs.harvard.edu/abs/2022MNRAS.517.4620R},
 archiveprefix = {arXiv},
 author = {{Rosenberg}, Erik and {Gratton}, Steven and {Efstathiou}, George},
 doi = {10.1093/mnras/stac2744},
 eprint = {2205.10869},
 journal = {\mnras},
 month = {December},
 number = {3},
 pages = {4620-4636},
 primaryclass = {astro-ph.CO},
 title = {{CMB power spectra and cosmological parameters from Planck PR4 with CamSpec}},
 volume = {517},
 year = {2022}
}

@ARTICLE{908974,
  author={Martinez, A.M. and Kak, A.C.},
  journal={IEEE Transactions on Pattern Analysis and Machine Intelligence}, 
  title={PCA versus LDA}, 
  year={2001},
  volume={23},
  number={2},
  pages={228-233},
  doi={10.1109/34.908974}}

@article{2411.06000,
 adsurl = {https://ui.adsabs.harvard.edu/abs/2025PhRvD.111h3534G},
 archiveprefix = {arXiv},
 author = {{Ge}, F. and {Millea}, M. and {Camphuis}, E. and {Daley}, C. and {Huang}, N. and {Omori}, Y. and {Quan}, W. and {Anderes}, E. and {Anderson}, A.~J. and {Ansarinejad}, B. and {Archipley}, M. and {Balkenhol}, L. and {Benabed}, K. and {Bender}, A.~N. and {Benson}, B.~A. and {Bianchini}, F. and {Bleem}, L.~E. and {Bouchet}, F.~R. and {Bryant}, L. and {Carlstrom}, J.~E. and {Chang}, C.~L. and {Chaubal}, P. and {Chen}, G. and {Chichura}, P.~M. and {Chokshi}, A. and {Chou}, T.-L. and {Coerver}, A. and {Crawford}, T.~M. and {de Haan}, T. and {Dibert}, K.~R. and {Dobbs}, M.~A. and {Doohan}, M. and {Doussot}, A. and {Dutcher}, D. and {Everett}, W. and {Feng}, C. and {Ferguson}, K.~R. and {Fichman}, K. and {Foster}, A. and {Galli}, S. and {Gambrel}, A.~E. and {Gardner}, R.~W. and {Goeckner-Wald}, N. and {Gualtieri}, R. and {Guidi}, F. and {Guns}, S. and {Halverson}, N.~W. and {Hivon}, E. and {Holder}, G.~P. and {Holzapfel}, W.~L. and {Hood}, J.~C. and {Howe}, D. and {Hryciuk}, A. and {K{\'e}ruzor{\'e}}, F. and {Khalife}, A.~R. and {Knox}, L. and {Korman}, M. and {Kornoelje}, K. and {Kuo}, C.-L. and {Lee}, A.~T. and {Levy}, K. and {Lowitz}, A.~E. and {Lu}, C. and {Maniyar}, A. and {Martsen}, E.~S. and {Menanteau}, F. and {Montgomery}, J. and {Nakato}, Y. and {Natoli}, T. and {Noble}, G.~I. and {Pan}, Z. and {Paschos}, P. and {Phadke}, K.~A. and {Pollak}, A.~W. and {Prabhu}, K. and {Rahimi}, M. and {Rahlin}, A. and {Reichardt}, C.~L. and {Riebel}, D. and {Rouble}, M. and {Ruhl}, J.~E. and {Schiappucci}, E. and {Sobrin}, J.~A. and {Stark}, A.~A. and {Stephen}, J. and {Tandoi}, C. and {Thorne}, B. and {Trendafilova}, C. and {Umilta}, C. and {Vieira}, J.~D. and {Vitrier}, A. and {Wan}, Y. and {Whitehorn}, N. and {Wu}, W.~L.~K. and {Young}, M.~R. and {Zebrowski}, J.~A. and {SPT-3G Collaboration}},
 doi = {10.1103/PhysRevD.111.083534},
 eid = {083534},
 eprint = {2411.06000},
 journal = {\prd},
 month = {April},
 number = {8},
 pages = {083534},
 primaryclass = {astro-ph.CO},
 title = {{Cosmology from CMB lensing and delensed EE power spectra using 2019{\textendash}2020 SPT-3G polarization data}},
 volume = {111},
 year = {2025}
}

@article{2506.20707,
 adsurl = {https://ui.adsabs.harvard.edu/abs/2025arXiv250620707C},
 archiveprefix = {arXiv},
 author = {{Camphuis}, E. and {Quan}, W. and {Balkenhol}, L. and {Khalife}, A.~R. and {Ge}, F. and {Guidi}, F. and {Huang}, N. and {Lynch}, G.~P. and {Omori}, Y. and {Trendafilova}, C. and {Anderson}, A.~J. and {Ansarinejad}, B. and {Archipley}, M. and {Barry}, P.~S. and {Benabed}, K. and {Bender}, A.~N. and {Benson}, B.~A. and {Bianchini}, F. and {Bleem}, L.~E. and {Bouchet}, F.~R. and {Bryant}, L. and {Campitiello}, M.~G. and {Carlstrom}, J.~E. and {Chang}, C.~L. and {Chaubal}, P. and {Chichura}, P.~M. and {Chokshi}, A. and {Chou}, T. -L. and {Coerver}, A. and {Crawford}, T.~M. and {Daley}, C. and {de Haan}, T. and {Dibert}, K.~R. and {Dobbs}, M.~A. and {Doohan}, M. and {Doussot}, A. and {Dutcher}, D. and {Everett}, W. and {Feng}, C. and {Ferguson}, K.~R. and {Fichman}, K. and {Foster}, A. and {Galli}, S. and {Gambrel}, A.~E. and {Gardner}, R.~W. and {Goeckner-Wald}, N. and {Gualtieri}, R. and {Guns}, S. and {Halverson}, N.~W. and {Hivon}, E. and {Holder}, G.~P. and {Holzapfel}, W.~L. and {Hood}, J.~C. and {Hryciuk}, A. and {K{\'e}ruzor{\'e}}, F. and {Knox}, L. and {Korman}, M. and {Kornoelje}, K. and {Kuo}, C. -L. and {Levy}, K. and {Lowitz}, A.~E. and {Lu}, C. and {Maniyar}, A. and {Martsen}, E.~S. and {Menanteau}, F. and {Millea}, M. and {Montgomery}, J. and {Nakato}, Y. and {Natoli}, T. and {Noble}, G.~I. and {Ouellette}, A. and {Pan}, Z. and {Paschos}, P. and {Phadke}, K.~A. and {Pollak}, A.~W. and {Prabhu}, K. and {Raghunathan}, S. and {Rahimi}, M. and {Rahlin}, A. and {Reichardt}, C.~L. and {Rouble}, M. and {Ruhl}, J.~E. and {Schiappucci}, E. and {Simpson}, A. and {Sobrin}, J.~A. and {Stark}, A.~A. and {Stephen}, J. and {Tandoi}, C. and {Thorne}, B. and {Umilta}, C. and {Vieira}, J.~D. and {Vitrier}, A. and {Wan}, Y. and {Whitehorn}, N. and {Wu}, W.~L.~K. and {Young}, M.~R. and {Zebrowski}, J.~A.},
 doi = {10.48550/arXiv.2506.20707},
 eid = {arXiv:2506.20707},
 eprint = {2506.20707},
 journal = {arXiv e-prints},
 month = {June},
 pages = {arXiv:2506.20707},
 primaryclass = {astro-ph.CO},
 title = {{SPT-3G D1: CMB temperature and polarization power spectra and cosmology from 2019 and 2020 observations of the SPT-3G Main field}},
 year = {2025}
}

@article{Abbott2022DES,
 adsurl = {https://ui.adsabs.harvard.edu/abs/2022PhRvD.105b3520A},
 archiveprefix = {arXiv},
 author = {{Abbott}, T.~M.~C. and {Aguena}, M. and {Alarcon}, A. and {Allam}, S. and {Alves}, O. and {Amon}, A. and {Andrade-Oliveira}, F. and {Annis}, J. and {Avila}, S. and {Bacon}, D. and {Baxter}, E. and {Bechtol}, K. and {Becker}, M.~R. and {Bernstein}, G.~M. and {Bhargava}, S. and {Birrer}, S. and {Blazek}, J. and {Brandao-Souza}, A. and {Bridle}, S.~L. and {Brooks}, D. and {Buckley-Geer}, E. and {Burke}, D.~L. and {Camacho}, H. and {Campos}, A. and {Carnero Rosell}, A. and {Carrasco Kind}, M. and {Carretero}, J. and {Castander}, F.~J. and {Cawthon}, R. and {Chang}, C. and {Chen}, A. and {Chen}, R. and {Choi}, A. and {Conselice}, C. and {Cordero}, J. and {Costanzi}, M. and {Crocce}, M. and {da Costa}, L.~N. and {da Silva Pereira}, M.~E. and {Davis}, C. and {Davis}, T.~M. and {De Vicente}, J. and {DeRose}, J. and {Desai}, S. and {Di Valentino}, E. and {Diehl}, H.~T. and {Dietrich}, J.~P. and {Dodelson}, S. and {Doel}, P. and {Doux}, C. and {Drlica-Wagner}, A. and {Eckert}, K. and {Eifler}, T.~F. and {Elsner}, F. and {Elvin-Poole}, J. and {Everett}, S. and {Evrard}, A.~E. and {Fang}, X. and {Farahi}, A. and {Fernandez}, E. and {Ferrero}, I. and {Fert{\'e}}, A. and {Fosalba}, P. and {Friedrich}, O. and {Frieman}, J. and {Garc{\'\i}a-Bellido}, J. and {Gatti}, M. and {Gaztanaga}, E. and {Gerdes}, D.~W. and {Giannantonio}, T. and {Giannini}, G. and {Gruen}, D. and {Gruendl}, R.~A. and {Gschwend}, J. and {Gutierrez}, G. and {Harrison}, I. and {Hartley}, W.~G. and {Herner}, K. and {Hinton}, S.~R. and {Hollowood}, D.~L. and {Honscheid}, K. and {Hoyle}, B. and {Huff}, E.~M. and {Huterer}, D. and {Jain}, B. and {James}, D.~J. and {Jarvis}, M. and {Jeffrey}, N. and {Jeltema}, T. and {Kovacs}, A. and {Krause}, E. and {Kron}, R. and {Kuehn}, K. and {Kuropatkin}, N. and {Lahav}, O. and {Leget}, P.-F. and {Lemos}, P. and {Liddle}, A.~R. and {Lidman}, C. and {Lima}, M. and {Lin}, H. and {MacCrann}, N. and {Maia}, M.~A.~G. and {Marshall}, J.~L. and {Martini}, P. and {McCullough}, J. and {Melchior}, P. and {Mena-Fern{\'a}ndez}, J. and {Menanteau}, F. and {Miquel}, R. and {Mohr}, J.~J. and {Morgan}, R. and {Muir}, J. and {Myles}, J. and {Nadathur}, S. and {Navarro-Alsina}, A. and {Nichol}, R.~C. and {Ogando}, R.~L.~C. and {Omori}, Y. and {Palmese}, A. and {Pandey}, S. and {Park}, Y. and {Paz-Chinch{\'o}n}, F. and {Petravick}, D. and {Pieres}, A. and {Plazas Malag{\'o}n}, A.~A. and {Porredon}, A. and {Prat}, J. and {Raveri}, M. and {Rodriguez-Monroy}, M. and {Rollins}, R.~P. and {Romer}, A.~K. and {Roodman}, A. and {Rosenfeld}, R. and {Ross}, A.~J. and {Rykoff}, E.~S. and {Samuroff}, S. and {S{\'a}nchez}, C. and {Sanchez}, E. and {Sanchez}, J. and {Sanchez Cid}, D. and {Scarpine}, V. and {Schubnell}, M. and {Scolnic}, D. and {Secco}, L.~F. and {Serrano}, S. and {Sevilla-Noarbe}, I. and {Sheldon}, E. and {Shin}, T. and {Smith}, M. and {Soares-Santos}, M. and {Suchyta}, E. and {Swanson}, M.~E.~C. and {Tabbutt}, M. and {Tarle}, G. and {Thomas}, D. and {To}, C. and {Troja}, A. and {Troxel}, M.~A. and {Tucker}, D.~L. and {Tutusaus}, I. and {Varga}, T.~N. and {Walker}, A.~R. and {Weaverdyck}, N. and {Wechsler}, R. and {Weller}, J. and {Yanny}, B. and {Yin}, B. and {Zhang}, Y. and {Zuntz}, J. and {DES Collaboration}},
 doi = {10.1103/PhysRevD.105.023520},
 eid = {023520},
 eprint = {2105.13549},
 journal = {\prd},
 month = {January},
 number = {2},
 pages = {023520},
 primaryclass = {astro-ph.CO},
 title = {{Dark Energy Survey Year 3 results: Cosmological constraints from galaxy clustering and weak lensing}},
 volume = {105},
 year = {2022}
}

@article{Abitbol2025SO,
 adsurl = {https://ui.adsabs.harvard.edu/abs/2025JCAP...08..034A},
 archiveprefix = {arXiv},
 author = {{Abitbol}, M. and {Abril-Cabezas}, I. and {Adachi}, S. and {Ade}, P. and {Adler}, A.~E. and {Agrawal}, P. and {Aguirre}, J. and {Ahmed}, Z. and {Aiola}, S. and {Alford}, T. and {Ali}, A. and {Alonso}, D. and {Alvarez}, M.~A. and {An}, R. and {Arnold}, K. and {Ashton}, P. and {Atkins}, Z. and {Austermann}, J. and {Azzoni}, S. and {Baccigalupi}, C. and {Baleato Lizancos}, A. and {Barron}, D. and {Barry}, P. and {Bartlett}, J. and {Battaglia}, N. and {Battye}, R. and {Baxter}, E. and {Bazarko}, A. and {Beall}, J.~A. and {Bean}, R. and {Beck}, D. and {Beckman}, S. and {Begin}, J. and {Beheshti}, A. and {Beringue}, B. and {Bhandarkar}, T. and {Bhimani}, S. and {Bianchini}, F. and {Biermann}, E. and {Biquard}, S. and {Bixler}, B. and {Boada}, S. and {Boettger}, D. and {Bolliet}, B. and {Bond}, J.~R. and {Borrill}, J. and {Borrow}, J. and {Braithwaite}, C. and {Brien}, T.~L.~R. and {Brown}, M.~L. and {Bruno}, S.~M. and {Bryan}, S. and {Bustos}, R. and {Cai}, H. and {Calabrese}, E. and {Calafut}, V. and {Carl}, F.~M. and {Carones}, A. and {Carron}, J. and {Challinor}, A. and {Chanial}, P. and {Chen}, N. and {Cheung}, K. and {Chiang}, B. and {Chinone}, Y. and {Chluba}, J. and {Cho}, H.~S. and {Choi}, S.~K. and {Chu}, M. and {Clancy}, J. and {Clark}, S.~E. and {Clarke}, P. and {Cleary}, J. and {Clements}, D.~L. and {Connors}, J. and {Contaldi}, C. and {Coppi}, G. and {Corbett}, L. and {Cothard}, N.~F. and {Coulton}, W. and {Crowley}, K.~D. and {Crowley}, K.~T. and {Cukierman}, A. and {D'Ewart}, J.~M. and {Dachlythra}, K. and {Datta}, R. and {Day-Weiss}, S. and {de Haan}, T. and {Devlin}, M. and {Di Mascolo}, L. and {Dicker}, S. and {Dober}, B. and {Doux}, C. and {Dow}, P. and {Doyle}, S. and {Duell}, C.~J. and {Duff}, S.~M. and {Duivenvoorden}, A.~J. and {Dunkley}, J. and {Dutcher}, D. and {D{\"u}nner}, R. and {Edenton}, M. and {El Bouhargani}, H. and {Errard}, J. and {Fabbian}, G. and {Fanfani}, V. and {Farren}, G.~S. and {Fergusson}, J. and {Ferraro}, S. and {Flauger}, R. and {Foster}, A. and {Freese}, K. and {Frisch}, J.~C. and {Frolov}, A. and {Fuller}, G. and {Galitzki}, N. and {Gallardo}, P.~A. and {Galvez Ghersi}, J.~T. and {Ganga}, K. and {Gao}, J. and {Garrido}, X. and {Gawiser}, E. and {Gerbino}, M. and {Gerras}, R. and {Giardiello}, S. and {Gill}, A. and {Gilles}, V. and {Giri}, U. and {Gleave}, E. and {Gluscevic}, V. and {Goeckner-Wald}, N. and {Golec}, J.~E. and {Gordon}, S. and {Gralla}, M. and {Gratton}, S. and {Green}, D. and {Groh}, J.~C. and {Groppi}, C. and {Guan}, Y. and {Gupta}, N. and {Gudmundsson}, J.~E. and {Hagstotz}, S. and {Hargrave}, P. and {Haridas}, S. and {Harrington}, K. and {Harrison}, I. and {Hasegawa}, M. and {Hasselfield}, M. and {Haynes}, V. and {Hazumi}, M. and {He}, A. and {Healy}, E. and {Henderson}, S.~W. and {Hensley}, B.~S. and {Hertig}, E. and {Herv{\'\i}as-Caimapo}, C. and {Higuchi}, M. and {Hill}, C.~A. and {Hill}, J.~C. and {Hilton}, G. and {Hilton}, M. and {Hincks}, A.~D. and {Hinshaw}, G. and {Hlo{\v{z}}ek}, R. and {Ho}, A.~Y.~Q. and {Ho}, S. and {Ho}, S.~P. and {Hoang}, T.~D. and {Hoh}, J. and {Hornecker}, E. and {Hornsby}, A.~L. and {Hotinli}, S.~C. and {Huang}, Z. and {Huber}, Z.~B. and {Hubmayr}, J. and {Huffenberger}, K. and {Hughes}, J.~P. and {Idicherian Lonappan}, A. and {Ikape}, M. and {Irwin}, K. and {Iuliano}, J. and {Jaffe}, A.~H. and {Jain}, B. and {Jense}, H.~T. and {Jeong}, O. and {Johnson}, A. and {Johnson}, B.~R. and {Johnson}, M. and {Jones}, M. and {Jost}, B. and {Kaneko}, D. and {Karpel}, E.~D. and {Kasai}, Y. and {Katayama}, N. and {Keating}, B. and {Keller}, B. and {Keskitalo}, R. and {Kim}, J. and {Kisner}, T. and {Kiuchi}, K.},
 doi = {10.1088/1475-7516/2025/08/034},
 eid = {034},
 eprint = {2503.00636},
 journal = {\jcap},
 month = {August},
 number = {8},
 pages = {034},
 primaryclass = {astro-ph.IM},
 title = {{The Simons Observatory: science goals and forecasts for the enhanced Large Aperture Telescope}},
 volume = {2025},
 year = {2025}
}

@article{Aghanim2020PlanckCMB,
 adsurl = {https://ui.adsabs.harvard.edu/abs/2020A&A...641A...6P},
 archiveprefix = {arXiv},
 author = {{Planck Collaboration} and {Aghanim}, N. and {Akrami}, Y. and {Ashdown}, M. and {Aumont}, J. and {Baccigalupi}, C. and {Ballardini}, M. and {Banday}, A.~J. and {Barreiro}, R.~B. and {Bartolo}, N. and {Basak}, S. and {Battye}, R. and {Benabed}, K. and {Bernard}, J.-P. and {Bersanelli}, M. and {Bielewicz}, P. and {Bock}, J.~J. and {Bond}, J.~R. and {Borrill}, J. and {Bouchet}, F.~R. and {Boulanger}, F. and {Bucher}, M. and {Burigana}, C. and {Butler}, R.~C. and {Calabrese}, E. and {Cardoso}, J.-F. and {Carron}, J. and {Challinor}, A. and {Chiang}, H.~C. and {Chluba}, J. and {Colombo}, L.~P.~L. and {Combet}, C. and {Contreras}, D. and {Crill}, B.~P. and {Cuttaia}, F. and {de Bernardis}, P. and {de Zotti}, G. and {Delabrouille}, J. and {Delouis}, J.-M. and {Di Valentino}, E. and {Diego}, J.~M. and {Dor{\'e}}, O. and {Douspis}, M. and {Ducout}, A. and {Dupac}, X. and {Dusini}, S. and {Efstathiou}, G. and {Elsner}, F. and {En{\ss}lin}, T.~A. and {Eriksen}, H.~K. and {Fantaye}, Y. and {Farhang}, M. and {Fergusson}, J. and {Fernandez-Cobos}, R. and {Finelli}, F. and {Forastieri}, F. and {Frailis}, M. and {Fraisse}, A.~A. and {Franceschi}, E. and {Frolov}, A. and {Galeotta}, S. and {Galli}, S. and {Ganga}, K. and {G{\'e}nova-Santos}, R.~T. and {Gerbino}, M. and {Ghosh}, T. and {Gonz{\'a}lez-Nuevo}, J. and {G{\'o}rski}, K.~M. and {Gratton}, S. and {Gruppuso}, A. and {Gudmundsson}, J.~E. and {Hamann}, J. and {Handley}, W. and {Hansen}, F.~K. and {Herranz}, D. and {Hildebrandt}, S.~R. and {Hivon}, E. and {Huang}, Z. and {Jaffe}, A.~H. and {Jones}, W.~C. and {Karakci}, A. and {Keih{\"a}nen}, E. and {Keskitalo}, R. and {Kiiveri}, K. and {Kim}, J. and {Kisner}, T.~S. and {Knox}, L. and {Krachmalnicoff}, N. and {Kunz}, M. and {Kurki-Suonio}, H. and {Lagache}, G. and {Lamarre}, J.-M. and {Lasenby}, A. and {Lattanzi}, M. and {Lawrence}, C.~R. and {Le Jeune}, M. and {Lemos}, P. and {Lesgourgues}, J. and {Levrier}, F. and {Lewis}, A. and {Liguori}, M. and {Lilje}, P.~B. and {Lilley}, M. and {Lindholm}, V. and {L{\'o}pez-Caniego}, M. and {Lubin}, P.~M. and {Ma}, Y.-Z. and {Mac{\'\i}as-P{\'e}rez}, J.~F. and {Maggio}, G. and {Maino}, D. and {Mandolesi}, N. and {Mangilli}, A. and {Marcos-Caballero}, A. and {Maris}, M. and {Martin}, P.~G. and {Martinelli}, M. and {Mart{\'\i}nez-Gonz{\'a}lez}, E. and {Matarrese}, S. and {Mauri}, N. and {McEwen}, J.~D. and {Meinhold}, P.~R. and {Melchiorri}, A. and {Mennella}, A. and {Migliaccio}, M. and {Millea}, M. and {Mitra}, S. and {Miville-Desch{\^e}nes}, M.-A. and {Molinari}, D. and {Montier}, L. and {Morgante}, G. and {Moss}, A. and {Natoli}, P. and {N{\o}rgaard-Nielsen}, H.~U. and {Pagano}, L. and {Paoletti}, D. and {Partridge}, B. and {Patanchon}, G. and {Peiris}, H.~V. and {Perrotta}, F. and {Pettorino}, V. and {Piacentini}, F. and {Polastri}, L. and {Polenta}, G. and {Puget}, J.-L. and {Rachen}, J.~P. and {Reinecke}, M. and {Remazeilles}, M. and {Renzi}, A. and {Rocha}, G. and {Rosset}, C. and {Roudier}, G. and {Rubi{\~n}o-Mart{\'\i}n}, J.~A. and {Ruiz-Granados}, B. and {Salvati}, L. and {Sandri}, M. and {Savelainen}, M. and {Scott}, D. and {Shellard}, E.~P.~S. and {Sirignano}, C. and {Sirri}, G. and {Spencer}, L.~D. and {Sunyaev}, R. and {Suur-Uski}, A.-S. and {Tauber}, J.~A. and {Tavagnacco}, D. and {Tenti}, M. and {Toffolatti}, L. and {Tomasi}, M. and {Trombetti}, T. and {Valenziano}, L. and {Valiviita}, J. and {Van Tent}, B. and {Vibert}, L. and {Vielva}, P. and {Villa}, F. and {Vittorio}, N. and {Wandelt}, B.~D. and {Wehus}, I.~K. and {White}, M. and {White}, S.~D.~M. and {Zacchei}, A. and {Zonca}, A.},
 doi = {10.1051/0004-6361/201833910},
 eid = {A6},
 eprint = {1807.06209},
 journal = {\aap},
 month = {September},
 pages = {A6},
 primaryclass = {astro-ph.CO},
 title = {{Planck 2018 results. VI. Cosmological parameters}},
 volume = {641},
 year = {2020}
}

@article{Aghanim2020PlanckLensing,
 adsurl = {https://ui.adsabs.harvard.edu/abs/2020A&A...641A...8P},
 archiveprefix = {arXiv},
 author = {{Planck Collaboration} and {Aghanim}, N. and {Akrami}, Y. and {Ashdown}, M. and {Aumont}, J. and {Baccigalupi}, C. and {Ballardini}, M. and {Banday}, A.~J. and {Barreiro}, R.~B. and {Bartolo}, N. and {Basak}, S. and {Benabed}, K. and {Bernard}, J.-P. and {Bersanelli}, M. and {Bielewicz}, P. and {Bock}, J.~J. and {Bond}, J.~R. and {Borrill}, J. and {Bouchet}, F.~R. and {Boulanger}, F. and {Bucher}, M. and {Burigana}, C. and {Calabrese}, E. and {Cardoso}, J.-F. and {Carron}, J. and {Challinor}, A. and {Chiang}, H.~C. and {Colombo}, L.~P.~L. and {Combet}, C. and {Crill}, B.~P. and {Cuttaia}, F. and {de Bernardis}, P. and {de Zotti}, G. and {Delabrouille}, J. and {Di Valentino}, E. and {Diego}, J.~M. and {Dor{\'e}}, O. and {Douspis}, M. and {Ducout}, A. and {Dupac}, X. and {Efstathiou}, G. and {Elsner}, F. and {En{\ss}lin}, T.~A. and {Eriksen}, H.~K. and {Fantaye}, Y. and {Fernandez-Cobos}, R. and {Finelli}, F. and {Forastieri}, F. and {Frailis}, M. and {Fraisse}, A.~A. and {Franceschi}, E. and {Frolov}, A. and {Galeotta}, S. and {Galli}, S. and {Ganga}, K. and {G{\'e}nova-Santos}, R.~T. and {Gerbino}, M. and {Ghosh}, T. and {Gonz{\'a}lez-Nuevo}, J. and {G{\'o}rski}, K.~M. and {Gratton}, S. and {Gruppuso}, A. and {Gudmundsson}, J.~E. and {Hamann}, J. and {Handley}, W. and {Hansen}, F.~K. and {Herranz}, D. and {Hivon}, E. and {Huang}, Z. and {Jaffe}, A.~H. and {Jones}, W.~C. and {Karakci}, A. and {Keih{\"a}nen}, E. and {Keskitalo}, R. and {Kiiveri}, K. and {Kim}, J. and {Knox}, L. and {Krachmalnicoff}, N. and {Kunz}, M. and {Kurki-Suonio}, H. and {Lagache}, G. and {Lamarre}, J.-M. and {Lasenby}, A. and {Lattanzi}, M. and {Lawrence}, C.~R. and {Le Jeune}, M. and {Levrier}, F. and {Lewis}, A. and {Liguori}, M. and {Lilje}, P.~B. and {Lindholm}, V. and {L{\'o}pez-Caniego}, M. and {Lubin}, P.~M. and {Ma}, Y.-Z. and {Mac{\'\i}as-P{\'e}rez}, J.~F. and {Maggio}, G. and {Maino}, D. and {Mandolesi}, N. and {Mangilli}, A. and {Marcos-Caballero}, A. and {Maris}, M. and {Martin}, P.~G. and {Mart{\'\i}nez-Gonz{\'a}lez}, E. and {Matarrese}, S. and {Mauri}, N. and {McEwen}, J.~D. and {Melchiorri}, A. and {Mennella}, A. and {Migliaccio}, M. and {Miville-Desch{\^e}nes}, M.-A. and {Molinari}, D. and {Moneti}, A. and {Montier}, L. and {Morgante}, G. and {Moss}, A. and {Natoli}, P. and {Pagano}, L. and {Paoletti}, D. and {Partridge}, B. and {Patanchon}, G. and {Perrotta}, F. and {Pettorino}, V. and {Piacentini}, F. and {Polastri}, L. and {Polenta}, G. and {Puget}, J.-L. and {Rachen}, J.~P. and {Reinecke}, M. and {Remazeilles}, M. and {Renzi}, A. and {Rocha}, G. and {Rosset}, C. and {Roudier}, G. and {Rubi{\~n}o-Mart{\'\i}n}, J.~A. and {Ruiz-Granados}, B. and {Salvati}, L. and {Sandri}, M. and {Savelainen}, M. and {Scott}, D. and {Sirignano}, C. and {Sunyaev}, R. and {Suur-Uski}, A.-S. and {Tauber}, J.~A. and {Tavagnacco}, D. and {Tenti}, M. and {Toffolatti}, L. and {Tomasi}, M. and {Trombetti}, T. and {Valiviita}, J. and {Van Tent}, B. and {Vielva}, P. and {Villa}, F. and {Vittorio}, N. and {Wandelt}, B.~D. and {Wehus}, I.~K. and {White}, M. and {White}, S.~D.~M. and {Zacchei}, A. and {Zonca}, A.},
 doi = {10.1051/0004-6361/201833886},
 eid = {A8},
 eprint = {1807.06210},
 journal = {\aap},
 month = {September},
 pages = {A8},
 primaryclass = {astro-ph.CO},
 title = {{Planck 2018 results. VIII. Gravitational lensing}},
 volume = {641},
 year = {2020}
}

@article{Amon2022ANonlinear,
 adsurl = {https://ui.adsabs.harvard.edu/abs/2022MNRAS.516.5355A},
 archiveprefix = {arXiv},
 author = {{Amon}, Alexandra and {Efstathiou}, George},
 doi = {10.1093/mnras/stac2429},
 eprint = {2206.11794},
 journal = {\mnras},
 month = {November},
 number = {4},
 pages = {5355-5366},
 primaryclass = {astro-ph.CO},
 title = {{A non-linear solution to the S$_{8}$ tension?}},
 volume = {516},
 year = {2022}
}

@article{Balkenhol2023SPTCMB,
 adsurl = {https://ui.adsabs.harvard.edu/abs/2023PhRvD.108b3510B},
 archiveprefix = {arXiv},
 author = {{Balkenhol}, L. and {Dutcher}, D. and {Spurio Mancini}, A. and {Doussot}, A. and {Benabed}, K. and {Galli}, S. and {Ade}, P.~A.~R. and {Anderson}, A.~J. and {Ansarinejad}, B. and {Archipley}, M. and {Bender}, A.~N. and {Benson}, B.~A. and {Bianchini}, F. and {Bleem}, L.~E. and {Bouchet}, F.~R. and {Bryant}, L. and {Camphuis}, E. and {Carlstrom}, J.~E. and {Cecil}, T.~W. and {Chang}, C.~L. and {Chaubal}, P. and {Chichura}, P.~M. and {Chou}, T.-L. and {Coerver}, A. and {Crawford}, T.~M. and {Cukierman}, A. and {Daley}, C. and {de Haan}, T. and {Dibert}, K.~R. and {Dobbs}, M.~A. and {Everett}, W. and {Feng}, C. and {Ferguson}, K.~R. and {Foster}, A. and {Gambrel}, A.~E. and {Gardner}, R.~W. and {Goeckner-Wald}, N. and {Gualtieri}, R. and {Guidi}, F. and {Guns}, S. and {Halverson}, N.~W. and {Hivon}, E. and {Holder}, G.~P. and {Holzapfel}, W.~L. and {Hood}, J.~C. and {Huang}, N. and {Knox}, L. and {Korman}, M. and {Kuo}, C.-L. and {Lee}, A.~T. and {Lowitz}, A.~E. and {Lu}, C. and {Millea}, M. and {Montgomery}, J. and {Nakato}, Y. and {Natoli}, T. and {Noble}, G.~I. and {Novosad}, V. and {Omori}, Y. and {Padin}, S. and {Pan}, Z. and {Paschos}, P. and {Prabhu}, K. and {Quan}, W. and {Rahimi}, M. and {Rahlin}, A. and {Reichardt}, C.~L. and {Rouble}, M. and {Ruhl}, J.~E. and {Schiappucci}, E. and {Smecher}, G. and {Sobrin}, J.~A. and {Stark}, A.~A. and {Stephen}, J. and {Suzuki}, A. and {Tandoi}, C. and {Thompson}, K.~L. and {Thorne}, B. and {Tucker}, C. and {Umilta}, C. and {Vieira}, J.~D. and {Wang}, G. and {Whitehorn}, N. and {Wu}, W.~L.~K. and {Yefremenko}, V. and {Young}, M.~R. and {Zebrowski}, J.~A. and {SPT-3G Collaboration}},
 doi = {10.1103/PhysRevD.108.023510},
 eid = {023510},
 eprint = {2212.05642},
 journal = {\prd},
 month = {July},
 number = {2},
 pages = {023510},
 primaryclass = {astro-ph.CO},
 title = {{Measurement of the CMB temperature power spectrum and constraints on cosmology from the SPT-3G 2018 TT, TE, and EE dataset}},
 volume = {108},
 year = {2023}
}

@article{Bigwood2024WeakLensing,
 adsurl = {https://ui.adsabs.harvard.edu/abs/2024MNRAS.534..655B},
 archiveprefix = {arXiv},
 author = {{Bigwood}, L. and {Amon}, A. and {Schneider}, A. and {Salcido}, J. and {McCarthy}, I.~G. and {Preston}, C. and {Sanchez}, D. and {Sijacki}, D. and {Schaan}, E. and {Ferraro}, S. and {Battaglia}, N. and {Chen}, A. and {Dodelson}, S. and {Roodman}, A. and {Pieres}, A. and {Fert{\'e}}, A. and {Alarcon}, A. and {Drlica-Wagner}, A. and {Choi}, A. and {Navarro-Alsina}, A. and {Campos}, A. and {Ross}, A.~J. and {Carnero Rosell}, A. and {Yin}, B. and {Yanny}, B. and {S{\'a}nchez}, C. and {Chang}, C. and {Davis}, C. and {Doux}, C. and {Gruen}, D. and {Rykoff}, E.~S. and {Huff}, E.~M. and {Sheldon}, E. and {Tarsitano}, F. and {Andrade-Oliveira}, F. and {Bernstein}, G.~M. and {Giannini}, G. and {Diehl}, H.~T. and {Huang}, H. and {Harrison}, I. and {Sevilla-Noarbe}, I. and {Tutusaus}, I. and {Elvin-Poole}, J. and {McCullough}, J. and {Zuntz}, J. and {Blazek}, J. and {DeRose}, J. and {Cordero}, J. and {Prat}, J. and {Myles}, J. and {Eckert}, K. and {Bechtol}, K. and {Herner}, K. and {Secco}, L.~F. and {Gatti}, M. and {Raveri}, M. and {Kind}, M. Carrasco and {Becker}, M.~R. and {Troxel}, M.~A. and {Jarvis}, M. and {MacCrann}, N. and {Friedrich}, O. and {Alves}, O. and {Leget}, P. -F. and {Chen}, R. and {Rollins}, R.~P. and {Wechsler}, R.~H. and {Gruendl}, R.~A. and {Cawthon}, R. and {Allam}, S. and {Bridle}, S.~L. and {Pandey}, S. and {Everett}, S. and {Shin}, T. and {Hartley}, W.~G. and {Fang}, X. and {Zhang}, Y. and {Aguena}, M. and {Annis}, J. and {Bacon}, D. and {Bertin}, E. and {Bocquet}, S. and {Brooks}, D. and {Carretero}, J. and {Castander}, F.~J. and {da Costa}, L.~N. and {Pereira}, M.~E.~S. and {De Vicente}, J. and {Desai}, S. and {Doel}, P. and {Ferrero}, I. and {Flaugher}, B. and {Frieman}, J. and {Garc{\'\i}a-Bellido}, J. and {Gaztanaga}, E. and {Gutierrez}, G. and {Hinton}, S.~R. and {Hollowood}, D.~L. and {Honscheid}, K. and {Huterer}, D. and {James}, D.~J. and {Kuehn}, K. and {Lahav}, O. and {Lee}, S. and {Marshall}, J.~L. and {Mena-Fern{\'a}ndez}, J. and {Miquel}, R. and {Muir}, J. and {Paterno}, M. and {Plazas Malag{\'o}n}, A.~A. and {Porredon}, A. and {Romer}, A.~K. and {Samuroff}, S. and {Sanchez}, E. and {Sanchez Cid}, D. and {Smith}, M. and {Soares-Santos}, M. and {Suchyta}, E. and {Swanson}, M.~E.~C. and {Tarle}, G. and {To}, C. and {Weaverdyck}, N. and {Weller}, J. and {Wiseman}, P. and {Yamamoto}, M.},
 doi = {10.1093/mnras/stae2100},
 eprint = {2404.06098},
 journal = {\mnras},
 month = {October},
 number = {1},
 pages = {655-682},
 primaryclass = {astro-ph.CO},
 title = {{Weak lensing combined with the kinetic Sunyaev-Zel'dovich effect: a study of baryonic feedback}},
 volume = {534},
 year = {2024}
}

@article{Bigwood2025TheKinetic,
 adsurl = {https://ui.adsabs.harvard.edu/abs/2025arXiv251015822B},
 archiveprefix = {arXiv},
 author = {{Bigwood}, Leah and {Yamamoto}, Masaya and {Siegel}, Jared and {Amon}, Alexandra and {McCarthy}, Ian G. and {Dave}, Romeel and {Salcido}, Jaime and {Schaller}, Matthieu and {Schaye}, Joop and {Yang}, Tianyi},
 eid = {arXiv:2510.15822},
 eprint = {2510.15822},
 journal = {arXiv e-prints},
 month = {October},
 pages = {arXiv:2510.15822},
 primaryclass = {astro-ph.CO},
 title = {{The kinetic Sunyaev Zeldovich effect as a benchmark for AGN feedback models in hydrodynamical simulations: insights from DESI + ACT}},
 year = {2025}
}

@article{Bulbul2024TheSRGeRosita,
 adsurl = {https://ui.adsabs.harvard.edu/abs/2024A&A...685A.106B},
 archiveprefix = {arXiv},
 author = {{Bulbul}, E. and {Liu}, A. and {Kluge}, M. and {Zhang}, X. and {Sanders}, J.~S. and {Bahar}, Y.~E. and {Ghirardini}, V. and {Artis}, E. and {Seppi}, R. and {Garrel}, C. and {Ramos-Ceja}, M.~E. and {Comparat}, J. and {Balzer}, F. and {B{\"o}ckmann}, K. and {Br{\"u}ggen}, M. and {Clerc}, N. and {Dennerl}, K. and {Dolag}, K. and {Freyberg}, M. and {Grandis}, S. and {Gruen}, D. and {Kleinebreil}, F. and {Krippendorf}, S. and {Lamer}, G. and {Merloni}, A. and {Migkas}, K. and {Nandra}, K. and {Pacaud}, F. and {Predehl}, P. and {Reiprich}, T.~H. and {Schrabback}, T. and {Veronica}, A. and {Weller}, J. and {Zelmer}, S.},
 doi = {10.1051/0004-6361/202348264},
 eid = {A106},
 eprint = {2402.08452},
 journal = {\aap},
 month = {May},
 pages = {A106},
 primaryclass = {astro-ph.CO},
 title = {{The SRG/eROSITA All-Sky Survey. The first catalog of galaxy clusters and groups in the Western Galactic Hemisphere}},
 volume = {685},
 year = {2024}
}

@article{Dalal2023HSC,
 adsurl = {https://ui.adsabs.harvard.edu/abs/2023PhRvD.108l3519D},
 archiveprefix = {arXiv},
 author = {{Dalal}, Roohi and {Li}, Xiangchong and {Nicola}, Andrina and {Zuntz}, Joe and {Strauss}, Michael A. and {Sugiyama}, Sunao and {Zhang}, Tianqing and {Rau}, Markus M. and {Mandelbaum}, Rachel and {Takada}, Masahiro and {More}, Surhud and {Miyatake}, Hironao and {Kannawadi}, Arun and {Shirasaki}, Masato and {Taniguchi}, Takanori and {Takahashi}, Ryuichi and {Osato}, Ken and {Hamana}, Takashi and {Oguri}, Masamune and {Nishizawa}, Atsushi J. and {Malag{\'o}n}, Andr{\'e}s A. Plazas and {Sunayama}, Tomomi and {Alonso}, David and {Slosar}, An{\v{z}}e and {Luo}, Wentao and {Armstrong}, Robert and {Bosch}, James and {Hsieh}, Bau-Ching and {Komiyama}, Yutaka and {Lupton}, Robert H. and {Lust}, Nate B. and {MacArthur}, Lauren A. and {Miyazaki}, Satoshi and {Murayama}, Hitoshi and {Nishimichi}, Takahiro and {Okura}, Yuki and {Price}, Paul A. and {Tait}, Philip J. and {Tanaka}, Masayuki and {Wang}, Shiang-Yu},
 doi = {10.1103/PhysRevD.108.123519},
 eid = {123519},
 eprint = {2304.00701},
 journal = {\prd},
 month = {December},
 number = {12},
 pages = {123519},
 primaryclass = {astro-ph.CO},
 title = {{Hyper Suprime-Cam Year 3 results: Cosmology from cosmic shear power spectra}},
 volume = {108},
 year = {2023}
}

@article{Dalal2025DecipheringBaryonic,
 adsurl = {https://ui.adsabs.harvard.edu/abs/2025arXiv250704476D},
 archiveprefix = {arXiv},
 author = {{Dalal}, Nihar and {To}, Chun-Hao and {Hirata}, Chris and {Hyeon-Shin}, Tae and {Hilton}, Matt and {Pandey}, Shivam and {Bond}, J. Richard},
 doi = {10.48550/arXiv.2507.04476},
 eid = {arXiv:2507.04476},
 eprint = {2507.04476},
 journal = {arXiv e-prints},
 month = {July},
 pages = {arXiv:2507.04476},
 primaryclass = {astro-ph.CO},
 title = {{Deciphering Baryonic Feedback from ACT tSZ Galaxy Clusters}},
 year = {2025}
}

@article{Doux2022DES,
 adsurl = {https://ui.adsabs.harvard.edu/abs/2022MNRAS.515.1942D},
 archiveprefix = {arXiv},
 author = {{Doux}, C. and {Jain}, B. and {Zeurcher}, D. and {Lee}, J. and {Fang}, X. and {Rosenfeld}, R. and {Amon}, A. and {Camacho}, H. and {Choi}, A. and {Secco}, L.~F. and {Blazek}, J. and {Chang}, C. and {Gatti}, M. and {Gaztanaga}, E. and {Jeffrey}, N. and {Raveri}, M. and {Samuroff}, S. and {Alarcon}, A. and {Alves}, O. and {Andrade-Oliveira}, F. and {Baxter}, E. and {Bechtol}, K. and {Becker}, M.~R. and {Bernstein}, G.~M. and {Campos}, A. and {Carnero Rosell}, A. and {Carrasco Kind}, M. and {Cawthon}, R. and {Chen}, R. and {Cordero}, J. and {Crocce}, M. and {Davis}, C. and {DeRose}, J. and {Dodelson}, S. and {Drlica-Wagner}, A. and {Eckert}, K. and {Eifler}, T.~F. and {Elsner}, F. and {Elvin-Poole}, J. and {Everett}, S. and {Fert{\'e}}, A. and {Fosalba}, P. and {Friedrich}, O. and {Giannini}, G. and {Gruen}, D. and {Gruendl}, R.~A. and {Harrison}, I. and {Hartley}, W.~G. and {Herner}, K. and {Huang}, H. and {Huff}, E.~M. and {Huterer}, D. and {Jarvis}, M. and {Krause}, E. and {Kuropatkin}, N. and {Leget}, P. -F. and {Lemos}, P. and {Liddle}, A.~R. and {MacCrann}, N. and {McCullough}, J. and {Muir}, J. and {Myles}, J. and {Navarro-Alsina}, A. and {Pandey}, S. and {Park}, Y. and {Porredon}, A. and {Prat}, J. and {Rodriguez-Monroy}, M. and {Rollins}, R.~P. and {Roodman}, A. and {Ross}, A.~J. and {Rykoff}, E.~S. and {S{\'a}nchez}, C. and {Sanchez}, J. and {Sevilla-Noarbe}, I. and {Sheldon}, E. and {Shin}, T. and {Troja}, A. and {Troxel}, M.~A. and {Tutusaus}, I. and {Varga}, T.~N. and {Weaverdyck}, N. and {Wechsler}, R.~H. and {Yanny}, B. and {Yin}, B. and {Zhang}, Y. and {Zuntz}, J. and {Abbott}, T.~M.~C. and {Aguena}, M. and {Allam}, S. and {Annis}, J. and {Bacon}, D. and {Bertin}, E. and {Bocquet}, S. and {Brooks}, D. and {Burke}, D.~L. and {Carretero}, J. and {Costanzi}, M. and {da Costa}, L.~N. and {Pereira}, M.~E.~S. and {De Vicente}, J. and {Desai}, S. and {Diehl}, H.~T. and {Doel}, P. and {Ferrero}, I. and {Flaugher}, B. and {Frieman}, J. and {Garc{\'\i}a-Bellido}, J. and {Gerdes}, D.~W. and {Giannantonio}, T. and {Gschwend}, J. and {Gutierrez}, G. and {Hinton}, S.~R. and {Hollowood}, D.~L. and {Honscheid}, K. and {James}, D.~J. and {Kim}, A.~G. and {Kuehn}, K. and {Lahav}, O. and {Marshall}, J.~L. and {Menanteau}, F. and {Miquel}, R. and {Morgan}, R. and {Ogando}, R.~L.~C. and {Palmese}, A. and {Paz-Chinch{\'o}n}, F. and {Pieres}, A. and {Plazas Malag{\'o}n}, A.~A. and {Reil}, K. and {Sanchez}, E. and {Scarpine}, V. and {Serrano}, S. and {Smith}, M. and {Suchyta}, E. and {Swanson}, M.~E.~C. and {Tarle}, G. and {Thomas}, D. and {To}, C. and {Weller}, J. and {DES Collaboration}},
 doi = {10.1093/mnras/stac1826},
 eprint = {2203.07128},
 journal = {\mnras},
 month = {September},
 number = {2},
 pages = {1942-1972},
 primaryclass = {astro-ph.CO},
 title = {{Dark energy survey year 3 results: cosmological constraints from the analysis of cosmic shear in harmonic space}},
 volume = {515},
 year = {2022}
}

@article{Doux2025GoingBeyond,
 adsurl = {https://ui.adsabs.harvard.edu/abs/2025arXiv250616434D},
 archiveprefix = {arXiv},
 author = {{Doux}, Cyrille and {Karwal}, Tanvi},
 doi = {10.48550/arXiv.2506.16434},
 eid = {arXiv:2506.16434},
 eprint = {2506.16434},
 journal = {arXiv e-prints},
 month = {June},
 pages = {arXiv:2506.16434},
 primaryclass = {astro-ph.CO},
 title = {{Going beyond $S_8$: fast inference of the matter power spectrum from weak-lensing surveys}},
 year = {2025}
}

@article{Farhadi2025MachineLearning,
 adsurl = {https://ui.adsabs.harvard.edu/abs/2025arXiv251104770F},
 archiveprefix = {arXiv},
 author = {{Farhadi Khouzani}, Farshid and {Shaw}, Abinash Kumar and {La Plante}, Paul and {Mustafa Shareef}, Bryar and {Gewali}, Laxmi},
 doi = {10.48550/arXiv.2511.04770},
 eid = {arXiv:2511.04770},
 eprint = {2511.04770},
 journal = {arXiv e-prints},
 month = {November},
 pages = {arXiv:2511.04770},
 primaryclass = {astro-ph.CO},
 title = {{Machine Learning-Driven Analysis of kSZ Maps to Predict CMB Optical Depth $τ$}},
 year = {2025}
}

@article{Foreman-Mackey2013Emcee,
 adsurl = {https://ui.adsabs.harvard.edu/abs/2013PASP..125..306F},
 archiveprefix = {arXiv},
 author = {{Foreman-Mackey}, Daniel and {Hogg}, David W. and {Lang}, Dustin and {Goodman}, Jonathan},
 doi = {10.1086/670067},
 eprint = {1202.3665},
 journal = {\pasp},
 month = {March},
 number = {925},
 pages = {306},
 primaryclass = {astro-ph.IM},
 title = {{emcee: The MCMC Hammer}},
 volume = {125},
 year = {2013}
}

@article{Foreman2023SubtractingThe,
 adsurl = {https://ui.adsabs.harvard.edu/abs/2023PhRvD.107h3502F},
 archiveprefix = {arXiv},
 author = {{Foreman}, Simon and {Hotinli}, Selim C. and {Madhavacheril}, Mathew S. and {van Engelen}, Alexander and {Kreisch}, Christina D.},
 doi = {10.1103/PhysRevD.107.083502},
 eid = {083502},
 eprint = {2209.03973},
 journal = {\prd},
 month = {April},
 number = {8},
 pages = {083502},
 primaryclass = {astro-ph.CO},
 title = {{Subtracting the kinetic Sunyaev-Zeldovich effect from the cosmic microwave background with surveys of large-scale structure}},
 volume = {107},
 year = {2023}
}

@article{Hadzhiyska2024EvidenceFor,
 adsurl = {https://ui.adsabs.harvard.edu/abs/2025PhRvD.112h3509H},
 archiveprefix = {arXiv},
 author = {{Hadzhiyska}, B. and {Ferraro}, S. and {Ried Guachalla}, B. and {Schaan}, E. and {Aguilar}, J. and {Ahlen}, S. and {Battaglia}, N. and {Bond}, J.~R. and {Brooks}, D. and {Calabrese}, E. and {Choi}, S.~K. and {Claybaugh}, T. and {Coulton}, W.~R. and {Dawson}, K. and {Devlin}, M. and {Dey}, B. and {Doel}, P. and {Duivenvoorden}, A.~J. and {Dunkley}, J. and {Farren}, G.~S. and {Font-Ribera}, A. and {Forero-Romero}, J.~E. and {Gallardo}, P.~A. and {Gazta{\~n}aga}, E. and {Gontcho Gontcho}, S. and {Gralla}, M. and {Le Guillou}, L. and {Gutierrez}, G. and {Guy}, J. and {Hill}, J.~C. and {Hlo{\v{z}}ek}, R. and {Honscheid}, K. and {Juneau}, S. and {Kehoe}, R. and {Kisner}, T. and {Kremin}, A. and {Landriau}, M. and {Liu}, R.~H. and {Louis}, T. and {MacCrann}, N. and {de Macorra}, A. and {Madhavacheril}, M. and {Manera}, M. and {Meisner}, A. and {Miquel}, R. and {Moodley}, K. and {Moustakas}, J. and {Mroczkowski}, T. and {Naess}, S. and {Newman}, J. and {Niemack}, M.~D. and {Niz}, G. and {Page}, L. and {Palanque-Delabrouille}, N. and {Partridge}, B. and {Percival}, W.~J. and {Prada}, F. and {Qu}, F.~J. and {Rossi}, G. and {Sanchez}, E. and {Schlegel}, D. and {Schubnell}, M. and {Sherwin}, B. and {Sehgal}, N. and {Seo}, H. and {Sif{\'o}n}, C. and {Spergel}, D. and {Sprayberry}, D. and {Staggs}, S. and {Tarl{\'e}}, G. and {Vargas}, C. and {Vavagiakis}, E.~M. and {Weaver}, B.~A. and {Wollack}, E.~J. and {Zhou}, R. and {Zou}, H.},
 doi = {10.1103/kclp-x5j1},
 eid = {083509},
 eprint = {2407.07152},
 journal = {\prd},
 month = {October},
 number = {8},
 pages = {083509},
 primaryclass = {astro-ph.CO},
 title = {{Evidence for large baryonic feedback at low and intermediate redshifts from kinematic Sunyaev-Zel'dovich observations with ACT and DESI photometric galaxies}},
 volume = {112},
 year = {2025}
}

@book{Hastie2009itz,
 author = {Hastie, Trevor and Tibshirani, Robert and Friedman, Jerome},
 doi = {10.1007/978-0-387-84858-7},
 isbn = {978-0-387-84857-0, 978-0-387-84858-7},
 publisher = {Springer},
 title = {{The Elements of Statistical Learning}},
 year = {2009}
}

@article{Henden2018TheFABLE,
 adsurl = {https://ui.adsabs.harvard.edu/abs/2018MNRAS.479.5385H},
 archiveprefix = {arXiv},
 author = {{Henden}, Nicholas A. and {Puchwein}, Ewald and {Shen}, Sijing and {Sijacki}, Debora},
 doi = {10.1093/mnras/sty1780},
 eprint = {1804.05064},
 journal = {\mnras},
 month = {October},
 number = {4},
 pages = {5385-5412},
 primaryclass = {astro-ph.GA},
 title = {{The FABLE simulations: a feedback model for galaxies, groups, and clusters}},
 volume = {479},
 year = {2018}
}

@article{Ho2009FindingThe,
 adsurl = {https://ui.adsabs.harvard.edu/abs/2009arXiv0903.2845H},
 archiveprefix = {arXiv},
 author = {{Ho}, Shirley and {Dedeo}, Simon and {Spergel}, David},
 doi = {10.48550/arXiv.0903.2845},
 eid = {arXiv:0903.2845},
 eprint = {0903.2845},
 journal = {arXiv e-prints},
 month = {March},
 pages = {arXiv:0903.2845},
 primaryclass = {astro-ph.CO},
 title = {{Finding the Missing Baryons Using CMB as a Backlight}},
 year = {2009}
}

@article{Karim2025DESIDR2,
 adsurl = {https://ui.adsabs.harvard.edu/abs/2025PhRvD.112h3515A},
 archiveprefix = {arXiv},
 author = {{Abdul Karim}, M. and {Aguilar}, J. and {Ahlen}, S. and {Alam}, S. and {Allen}, L. and {Prieto}, C. Allende and {Alves}, O. and {Anand}, A. and {Andrade}, U. and {Armengaud}, E. and {Aviles}, A. and {Bailey}, S. and {Baltay}, C. and {Bansal}, P. and {Bault}, A. and {Behera}, J. and {BenZvi}, S. and {Bianchi}, D. and {Blake}, C. and {Brieden}, S. and {Brodzeller}, A. and {Brooks}, D. and {Buckley-Geer}, E. and {Burtin}, E. and {Calderon}, R. and {Canning}, R. and {Rosell}, A. Carnero and {Carrilho}, P. and {Casas}, L. and {Castander}, F.~J. and {Charles}, M. and {Chaussidon}, E. and {Chaves-Montero}, J. and {Chebat}, D. and {Chen}, X. and {Claybaugh}, T. and {Cole}, S. and {Cooper}, A.~P. and {Cuceu}, A. and {Dawson}, K.~S. and {de la Macorra}, A. and {de Mattia}, A. and {Deiosso}, N. and {Della Costa}, J. and {Demina}, R. and {Dey}, A. and {Dey}, B. and {Ding}, Z. and {Doel}, P. and {Edelstein}, J. and {Eisenstein}, D.~J. and {Elbers}, W. and {Fagrelius}, P. and {Fanning}, K. and {Fern{\'a}ndez-Garc{\'\i}a}, E. and {Ferraro}, S. and {Font-Ribera}, A. and {Forero-Romero}, J.~E. and {Frenk}, C.~S. and {Garcia-Quintero}, C. and {Garrison}, L.~H. and {Gazta{\~n}aga}, E. and {Gil-Mar{\'\i}n}, H. and {Gontcho A Gontcho}, S. and {Gonzalez}, D. and {Gonzalez-Morales}, A.~X. and {Gordon}, C. and {Green}, D. and {Gutierrez}, G. and {Guy}, J. and {Hadzhiyska}, B. and {Hahn}, C. and {He}, S. and {Herbold}, M. and {Herrera-Alcantar}, H.~K. and {Ho}, M.-F. and {Honscheid}, K. and {Howlett}, C. and {Huterer}, D. and {Ishak}, M. and {Juneau}, S. and {Kamble}, N.~V. and {Kara{\c{c}}ayl{\i}}, N.~G. and {Kehoe}, R. and {Kent}, S. and {Kim}, A.~G. and {Kirkby}, D. and {Kisner}, T. and {Koposov}, S.~E. and {Kremin}, A. and {Krolewski}, A. and {Lahav}, O. and {Lamman}, C. and {Landriau}, M. and {Lang}, D. and {Lasker}, J. and {Le Goff}, J.~M. and {Le Guillou}, L. and {Leauthaud}, A. and {Levi}, M.~E. and {Li}, Q. and {Li}, T.~S. and {Lodha}, K. and {Lokken}, M. and {Lozano-Rodr{\'\i}guez}, F. and {Magneville}, C. and {Manera}, M. and {Martini}, P. and {Matthewson}, W.~L. and {Meisner}, A. and {Mena-Fern{\'a}ndez}, J. and {Menegas}, A. and {Mergulh{\~a}o}, T. and {Miquel}, R. and {Moustakas}, J. and {Mu{\~n}oz-Guti{\'e}rrez}, A. and {Mu{\~n}oz-Santos}, D. and {Myers}, A.~D. and {Nadathur}, S. and {Naidoo}, K. and {Napolitano}, L. and {Newman}, J.~A. and {Niz}, G. and {Noriega}, H.~E. and {Paillas}, E. and {Palanque-Delabrouille}, N. and {Pan}, J. and {Peacock}, J.~A. and {Pellejero Ibanez}, M. and {Percival}, W.~J. and {P{\'e}rez-Fern{\'a}ndez}, A. and {P{\'e}rez-R{\`a}fols}, I. and {Pieri}, M.~M. and {Poppett}, C. and {Prada}, F. and {Rabinowitz}, D. and {Raichoor}, A. and {Ram{\'\i}rez-P{\'e}rez}, C. and {Rashkovetskyi}, M. and {Ravoux}, C. and {Rich}, J. and {Rocher}, A. and {Rockosi}, C. and {Rohlf}, J. and {Rom{\'a}n-Herrera}, J.~O. and {Ross}, A.~J. and {Rossi}, G. and {Ruggeri}, R. and {Ruhlmann-Kleider}, V. and {Samushia}, L. and {Sanchez}, E. and {Sanders}, N. and {Schlegel}, D. and {Schubnell}, M. and {Seo}, H. and {Shafieloo}, A. and {Sharples}, R. and {Silber}, J. and {Sinigaglia}, F. and {Sprayberry}, D. and {Tan}, T. and {Tarl{\'e}}, G. and {Taylor}, P. and {Turner}, W. and {Ure{\~n}a-L{\'o}pez}, L.~A. and {Vaisakh}, R. and {Valdes}, F. and {Valogiannis}, G. and {Vargas-Maga{\~n}a}, M. and {Verde}, L. and {Walther}, M. and {Weaver}, B.~A. and {Weinberg}, D.~H. and {White}, M. and {Wolfson}, M. and {Y{\`e}che}, C. and {Yu}, J. and {Zaborowski}, E.~A. and {Zarrouk}, P. and {Zhai}, Z. and {Zhang}, H. and {Zhao}, C. and {Zhao}, G.~B. and {Zhou}, R. and {Zou}, H. and {DESI Collaboration}},
 doi = {10.1103/tr6y-kpc6},
 eid = {083515},
 eprint = {2503.14738},
 journal = {\prd},
 month = {October},
 number = {8},
 pages = {083515},
 primaryclass = {astro-ph.CO},
 title = {{DESI DR2 results. II. Measurements of baryon acoustic oscillations and cosmological constraints}},
 volume = {112},
 year = {2025}
}

@article{Kovac2025BaryonificationII,
 adsurl = {https://ui.adsabs.harvard.edu/abs/2025JCAP...11..046K},
 archiveprefix = {arXiv},
 author = {{Kova{\v{c}}}, Michael and {Nicola}, Andrina and {Bucko}, Jozef and {Schneider}, Aurel and {Reischke}, Robert and {Giri}, Sambit K. and {Teyssier}, Romain and {Schaller}, Matthieu and {Schaye}, Joop},
 doi = {10.1088/1475-7516/2025/11/046},
 eid = {046},
 eprint = {2507.07991},
 journal = {\jcap},
 month = {November},
 number = {11},
 pages = {046},
 primaryclass = {astro-ph.CO},
 title = {{Baryonification II: constraining feedback with X-ray and kinematic Sunyaev-Zel'dovich observations}},
 volume = {2025},
 year = {2025}
}

@article{Kugel2023FLAMINGOCalibrating,
 adsurl = {https://ui.adsabs.harvard.edu/abs/2023MNRAS.526.6103K},
 archiveprefix = {arXiv},
 author = {{Kugel}, Roi and {Schaye}, Joop and {Schaller}, Matthieu and {Helly}, John C. and {Braspenning}, Joey and {Elbers}, Willem and {Frenk}, Carlos S. and {McCarthy}, Ian G. and {Kwan}, Juliana and {Salcido}, Jaime and {van Daalen}, Marcel P. and {Vandenbroucke}, Bert and {Bah{\'e}}, Yannick M. and {Borrow}, Josh and {Chaikin}, Evgenii and {Hu{\v{s}}ko}, Filip and {Jenkins}, Adrian and {Lacey}, Cedric G. and {Nobels}, Folkert S.~J. and {Vernon}, Ian},
 doi = {10.1093/mnras/stad2540},
 eprint = {2306.05492},
 journal = {\mnras},
 month = {December},
 number = {4},
 pages = {6103-6127},
 primaryclass = {astro-ph.CO},
 title = {{FLAMINGO: calibrating large cosmological hydrodynamical simulations with machine learning}},
 volume = {526},
 year = {2023}
}

@software{Lewis2011CAMB,
 adsurl = {https://ui.adsabs.harvard.edu/abs/2011ascl.soft02026L},
 archiveprefix = {ascl},
 author = {{Lewis}, Antony and {Challinor}, Anthony},
 eid = {ascl:1102.026},
 eprint = {1102.026},
 howpublished = {Astrophysics Source Code Library, record ascl:1102.026},
 month = {February},
 title = {{CAMB: Code for Anisotropies in the Microwave Background}},
 year = {2011}
}

@article{Li2023HSC,
 adsurl = {https://ui.adsabs.harvard.edu/abs/2023PhRvD.108l3518L},
 archiveprefix = {arXiv},
 author = {{Li}, Xiangchong and {Zhang}, Tianqing and {Sugiyama}, Sunao and {Dalal}, Roohi and {Terasawa}, Ryo and {Rau}, Markus M. and {Mandelbaum}, Rachel and {Takada}, Masahiro and {More}, Surhud and {Strauss}, Michael A. and {Miyatake}, Hironao and {Shirasaki}, Masato and {Hamana}, Takashi and {Oguri}, Masamune and {Luo}, Wentao and {Nishizawa}, Atsushi J. and {Takahashi}, Ryuichi and {Nicola}, Andrina and {Osato}, Ken and {Kannawadi}, Arun and {Sunayama}, Tomomi and {Armstrong}, Robert and {Bosch}, James and {Komiyama}, Yutaka and {Lupton}, Robert H. and {Lust}, Nate B. and {MacArthur}, Lauren A. and {Miyazaki}, Satoshi and {Murayama}, Hitoshi and {Nishimichi}, Takahiro and {Okura}, Yuki and {Price}, Paul A. and {Tait}, Philip J. and {Tanaka}, Masayuki and {Wang}, Shiang-Yu},
 doi = {10.1103/PhysRevD.108.123518},
 eid = {123518},
 eprint = {2304.00702},
 journal = {\prd},
 month = {December},
 number = {12},
 pages = {123518},
 primaryclass = {astro-ph.CO},
 title = {{Hyper Suprime-Cam Year 3 results: Cosmology from cosmic shear two-point correlation functions}},
 volume = {108},
 year = {2023}
}

@article{McCarthy2017TheBAHAMAS,
 adsurl = {https://ui.adsabs.harvard.edu/abs/2017MNRAS.465.2936M},
 archiveprefix = {arXiv},
 author = {{McCarthy}, Ian G. and {Schaye}, Joop and {Bird}, Simeon and {Le Brun}, Amandine M.~C.},
 doi = {10.1093/mnras/stw2792},
 eprint = {1603.02702},
 journal = {\mnras},
 month = {March},
 number = {3},
 pages = {2936-2965},
 primaryclass = {astro-ph.CO},
 title = {{The BAHAMAS project: calibrated hydrodynamical simulations for large-scale structure cosmology}},
 volume = {465},
 year = {2017}
}

@article{McCarthy2025FlamingoCombining,
 adsurl = {https://ui.adsabs.harvard.edu/abs/2025MNRAS.540..143M},
 archiveprefix = {arXiv},
 author = {{McCarthy}, Ian G. and {Amon}, Alexandra and {Schaye}, Joop and {Schaan}, Emmanuel and {Angulo}, Raul E. and {Salcido}, Jaime and {Schaller}, Matthieu and {Bigwood}, Leah and {Elbers}, Willem and {Kugel}, Roi and {Helly}, John C. and {Forouhar Moreno}, Victor J. and {Frenk}, Carlos S. and {McGibbon}, Robert J. and {Ondaro-Mallea}, Lurdes and {van Daalen}, Marcel P.},
 doi = {10.1093/mnras/staf731},
 eprint = {2410.19905},
 journal = {\mnras},
 month = {June},
 number = {1},
 pages = {143-163},
 primaryclass = {astro-ph.CO},
 title = {{FLAMINGO: combining kinetic SZ effect and galaxy{\textendash}galaxy lensing measurements to gauge the impact of feedback on large-scale structure}},
 volume = {540},
 year = {2025}
}

@article{Mead2021HMCODE,
 adsurl = {https://ui.adsabs.harvard.edu/abs/2021MNRAS.502.1401M},
 archiveprefix = {arXiv},
 author = {{Mead}, A.~J. and {Brieden}, S. and {Tr{\"o}ster}, T. and {Heymans}, C.},
 doi = {10.1093/mnras/stab082},
 eprint = {2009.01858},
 journal = {\mnras},
 month = {March},
 number = {1},
 pages = {1401-1422},
 primaryclass = {astro-ph.CO},
 title = {{HMCODE-2020: improved modelling of non-linear cosmological power spectra with baryonic feedback}},
 volume = {502},
 year = {2021}
}

@article{Mootoovaloo2022KernelBased,
 adsurl = {https://ui.adsabs.harvard.edu/abs/2022A&C....3800508M},
 archiveprefix = {arXiv},
 author = {{Mootoovaloo}, A. and {Jaffe}, A.~H. and {Heavens}, A.~F. and {Leclercq}, F.},
 doi = {10.1016/j.ascom.2021.100508},
 eid = {100508},
 eprint = {2105.02256},
 journal = {Astronomy and Computing},
 month = {January},
 pages = {100508},
 primaryclass = {astro-ph.CO},
 title = {{Kernel-based emulator for the 3D matter power spectrum from CLASS}},
 volume = {38},
 year = {2022}
}

@article{Padmanabhan2012A2Percent,
 adsurl = {https://ui.adsabs.harvard.edu/abs/2012MNRAS.427.2132P},
 archiveprefix = {arXiv},
 author = {{Padmanabhan}, Nikhil and {Xu}, Xiaoying and {Eisenstein}, Daniel J. and {Scalzo}, Richard and {Cuesta}, Antonio J. and {Mehta}, Kushal T. and {Kazin}, Eyal},
 doi = {10.1111/j.1365-2966.2012.21888.x},
 eprint = {1202.0090},
 journal = {\mnras},
 month = {December},
 number = {3},
 pages = {2132-2145},
 primaryclass = {astro-ph.CO},
 title = {{A 2 per cent distance to z = 0.35 by reconstructing baryon acoustic oscillations - I. Methods and application to the Sloan Digital Sky Survey}},
 volume = {427},
 year = {2012}
}

@article{Perez2025ReconstructingThe,
 adsurl = {https://ui.adsabs.harvard.edu/abs/2025PhRvD.112f3510P},
 archiveprefix = {arXiv},
 author = {{Perez Sarmiento}, Karen and {Lagu{\"e}}, Alex and {Madhavacheril}, Mathew S. and {Jain}, Bhuvnesh and {Sherwin}, Blake},
 doi = {10.1103/bzrj-76sn},
 eid = {063510},
 eprint = {2502.06687},
 journal = {\prd},
 month = {September},
 number = {6},
 pages = {063510},
 primaryclass = {astro-ph.CO},
 title = {{Reconstructing the shape of the nonlinear matter power spectrum using CMB lensing and cosmic shear}},
 volume = {112},
 year = {2025}
}



\appendix
\section{Imperfect velocity reconstruction}
\label{app:systematics}

Throughout this work, we have assumed that the galaxy velocities were known. In practice, however, the galaxy velocities are estimated by solving the continuity equation~\citep{Padmanabhan2012A2Percent}. The resulting estimated radial velocity field (which we will denote $\hat v_r$) is strongly correlated with the true velocity field on large scales. Taking the large-scale limit, we can approximate 
\begin{align}
    P_{\hat v_r v_r}(k, \mu, z) &\approx r^2_v P_{ v_r v_r}(k, \mu, z),\;\;\;k\ll 1.
\end{align}
where $r^2_v$ is a constant capturing the correlation between the reconstructed and true velocities. It can be estimated for any given galaxy survey using simulations.

\section{Training with limited information}\label{app:info}
\begin{figure}
    \centering
    \begin{subfigure}{\linewidth}
        \centering
        \includegraphics[width=0.875\linewidth]{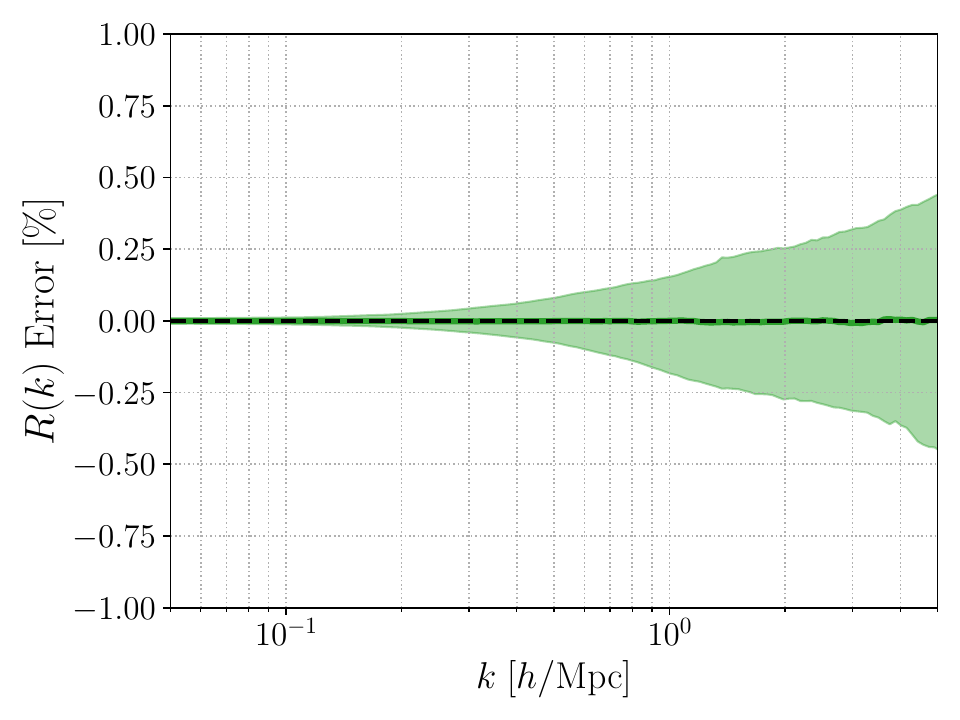}
        \caption{}
    \end{subfigure}
    
    \begin{subfigure}{\linewidth}
        \centering
        \includegraphics[width=0.875\linewidth]{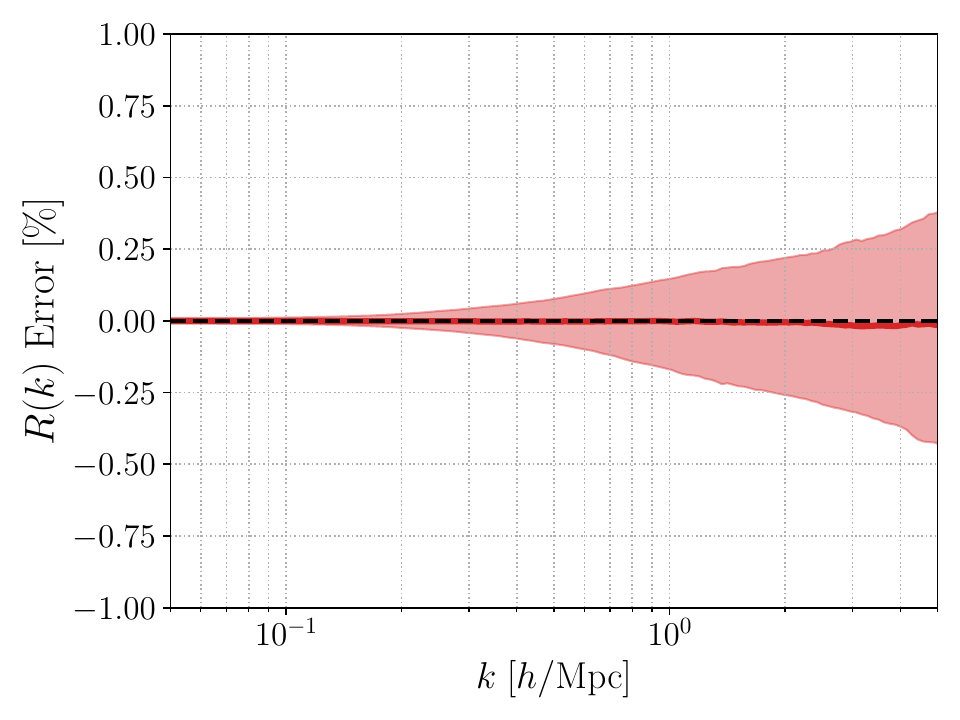}
        \caption{}
    \end{subfigure}

    \begin{subfigure}{\linewidth}
        \centering
        \includegraphics[width=0.875\linewidth]{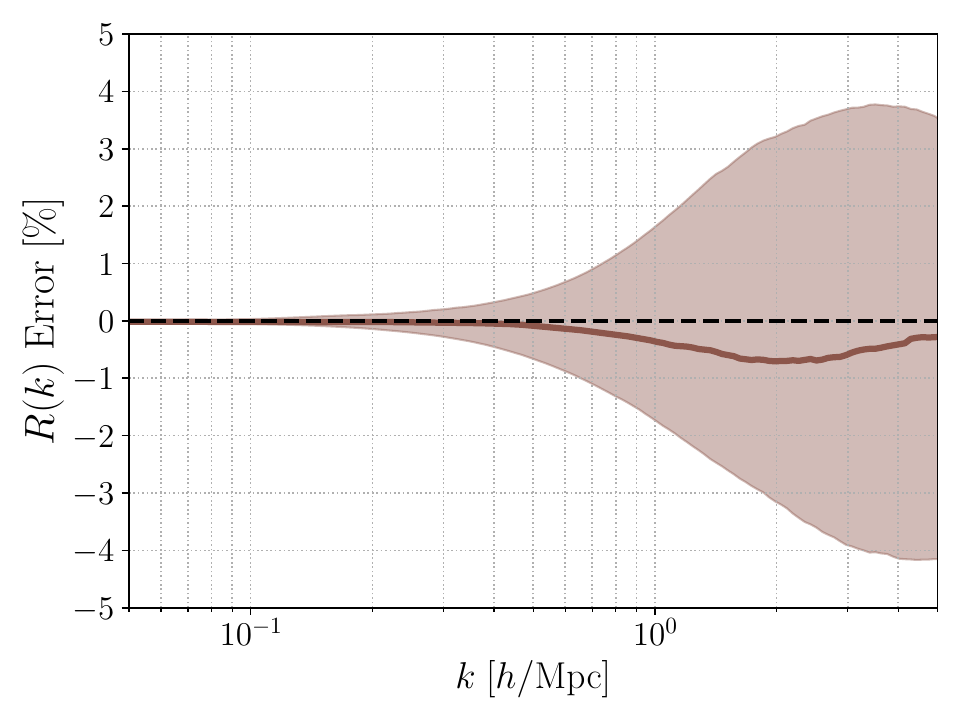}
        \caption{}
    \end{subfigure}
    \caption{(\textit{Top}) Test set validation after training the emulator without any of the $a_n^{gg}$ parameters. (\textit{Center}) Test set validation after training the emulator with only $a_0^{gg}$ for the $a_n^{gg}$ parameters. (\textit{Bottom}) Test set validation after training the emulator without any of the $a_n^{\hat bb}$ parameters. The range in the bottom panel was extended to capture the full range of deviations. \label{fig:hod-no-galaxies}}
\end{figure}

In this appendix, we test the capacity of the emulator approach to infer the shape of the matter power spectrum without being given the full galaxy angular power spectrum. Most of the small-scale matter power spectrum information is contained in the Doppler $b$ cross-spectrum. In Fig.~\ref{fig:hod-no-galaxies}, we test the network's performance without the galaxy power spectrum observations. We use the same technique as previously, retaining 5\% of the data as a validation set. In the first case, we remove the four parameters used to fit $C_\ell^{gg}$, and in the second case, we keep the constant parameter of the fit. This constant acts as a proxy for the number density as it is proportional to the galaxy shot noise. 

We observe a degradation in network performance compared to the case with all nine features, as shown in Fig.~\ref{fig:hod_residuals}. Including the shot noise does not visually change the quality of the fit, but it slightly improves the mean absolute error of the prediction on the test set. While most of the neural network predictions remain within 1\%, we observe a significant number of catastrophic failures, which are absent when including the information from the full galaxy auto-spectrum (see Fig.~\ref{fig:hod_residuals}). As predicted, the neural network fails when trained only using the galaxy auto-spectrum and the redshift. This demonstrates that the network does not learn the shape of the matter power spectrum from the galaxies themselves, but from the small-scale information contained in Doppler $b$ maps. This test ensures that our network does not over-fit the galaxy spectrum and that there is no information leakage in our pipeline.

We conclude from this test that the galaxy auto-spectrum contains valuable information, making the emulator more adaptable to different galaxy populations. Adding other observables, such as gravitational lensing, thermal Sunyaev-Zeldovich, or X-ray, to the input vector could also contribute to the accuracy of the emulator. Given our pre-processing steps, our training process is easily adaptable to higher-dimensional data.

\end{document}